\documentclass[11pt]{article}
\usepackage{epsfig,bm,epsf,latexsym, amsmath, amssymb, bbm, epsf, graphicx}
\usepackage{cjhebrew}
\usepackage[left=.75in,top=.75in,right=.75in,nohead]{geometry}
\usepackage[small]{caption}
\usepackage{setspace}

\newcommand{\bc}{{\bf c}}
\newcommand{\be}{{\hat{\bf e}}}
\newcommand{\bn}{{\bf n}}
\newcommand{\bs}{{\bf s}}

\newcommand{\bx}{{\bf x}}

\newcommand{\bI}{{\bf I}}
\newcommand{\Li}{{\rm Li}}
\newcommand{\bbsigma}{{\bm \sigma}}

\newcommand{\obj}{{\text{\cjRL{m|}}}}
\newcommand{\oobj}{{\text{\cjRL{M}}}}

\begin{document}

\title{\bf Exact feature probabilities in images with occlusion}
\author{Xaq Pitkow\\
\footnotesize Center for Theoretical Neuroscience, Columbia University}
\date{\footnotesize \today}
\maketitle

\begin{abstract}
To understand the computations of our visual system, it is important to understand also the natural environment it evolved to interpret. Unfortunately, existing models of the visual environment are either unrealistic or too complex for mathematical description. Here we describe a naturalistic image model and present a mathematical solution for the statistical relationships between the image features and model variables. The world described by this model is composed of independent, opaque, textured objects which occlude each other. This simple structure allows us to calculate the joint probability distribution of image values sampled at multiple arbitrarily located points, without approximation. This result can be converted into probabilistic relationships between observable image features as well as between the unobservable properties that caused these features, including object boundaries and relative depth. Using these results we explain the causes of a wide range of natural scene properties, including highly non-gaussian distributions of image features and causal relations between pairs of edges. We discuss the implications of this description of natural scenes for the study of vision.
\end{abstract}


\section{Introduction}

A major goal of vision is to identify physical objects in the world, and their attributes. The relevant sensory evidence --- an image --- is ambiguous. A visual system must make guesses to interpret this sensory information, and good guesses should account for the statistics of the input. Consequently, the statistical structure of natural images has become a subject of fundamental importance for applications ranging from computer graphics to neuroscience: Understanding and exploiting natural regularities should lead to better visual performance and improved visual representations, whether in image compression or in the brain.

Most previous studies of natural scene statistics have characterized natural scenes as linear superpositions of image features. Principal Components Analysis \cite{Hancock:1992p2760,Liu:1994p2786}, Independent Components Analysis \cite{Olshausen:1996p2908,Bell:1997p2824}, and wavelet transforms \cite{Portilla:2000p5348} each identify related sets of features that when added together can efficiently reconstitute a natural image. Other methods have improved upon these purely linear, additive descriptions by including multiplicative modulation (e.g. Gaussian scale mixtures \cite{Wainwright:2000p4246}, hierarchically correlated variances \cite{Karklin:2005p4388}). These nonlinear enhancements are useful for representing textures, where common variables like surface properties and illumination intensity and direction naturally co-modulate the contrast of features like local orientation. Yet because visual images are not caused by summation but by occlusion, it is important to develop models of natural images that are constructed using more accurate nonlinear combinations of features \cite{Reinagel:2001p2570,Simoncelli:2001p4288}.

\begin{figure}
\centering
\includegraphics*[width=7in]{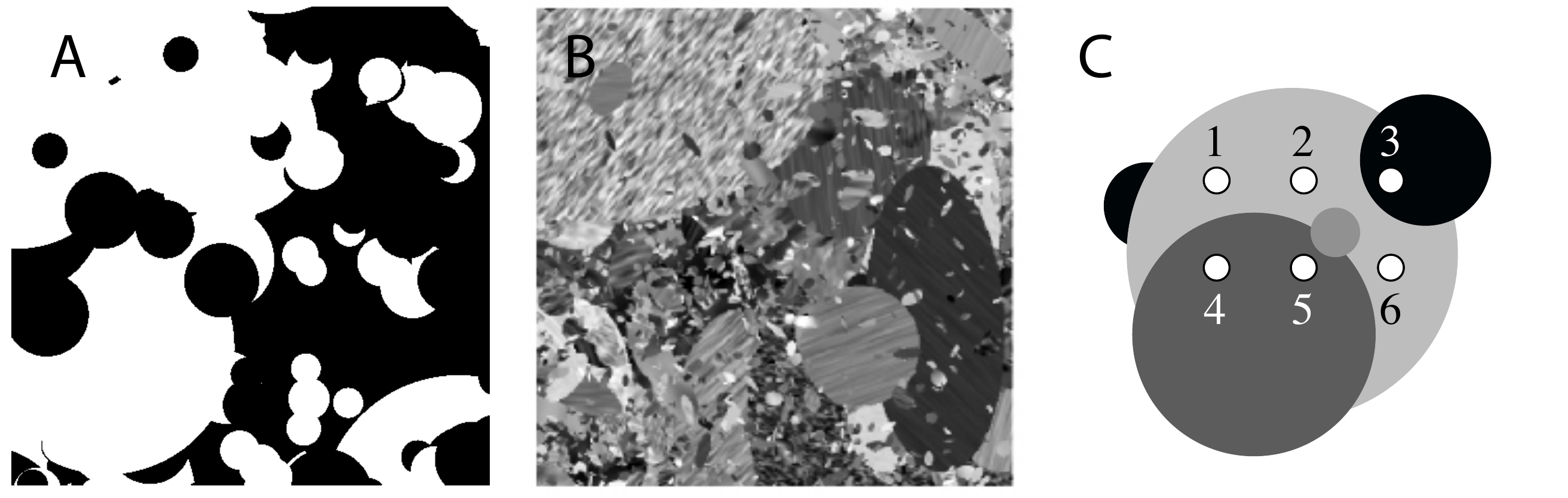}
\caption{Example images generated by the dead leaves model. We see layers of objects with random sizes, shapes, colors and positions that occlude other objects below. ({\bf A}) All objects are black or white circles with a relatively narrow range of sizes. ({\bf B}) All objects are textured ellipses with a broad range of sizes drawn from a distribution proportional to ${\rm size}^{-3}$, producing approximate scale invariance \cite{Ruderman:1997p825, Lee:2001p1}. Straightforward generalizations allow other ensembles of shape and texture. ({\bf C}) Illustration of an object membership function $\obj$. Pixels within a member set of $\obj$ all sample from the same object. Shown is an example dead leaves image with several objects (grey circles) and a set of six pixel locations (numbered points). For this configuration, the object membership function is $\obj=\{126|3|45\}$.}
\label{exampleDeadLeavesImage}
\end{figure}

We therefore chose a simplified model of natural images, colorfully known as the dead leaves model \cite{Matheron:1975p820}, for which occlusion is fundamental: The virtual world described by this model is composed of an infinitely deep stack of randomly positioned, flat objects (`leaves') that occlude each other (Figure \ref{exampleDeadLeavesImage}A--B). These objects have attributes of size, shape, texture, and color, independently drawn from specified distributions. While the model is only an approximation to our true physical environment, nonetheless it generates images that share many important attributes with natural images, most obviously the ubiquitous boundaries between relatively homogeneous regions \cite{Ruderman:1997p825}. It reproduces several known statistical properties of natural images, including the spatial power spectrum \cite{Ruderman:1997p825} and bivariate distributions of pixel intensities and wavelets coefficients \cite{Lee:2001p1}. Despite the demonstrated utility of the model, until now there has been no way to calculate higher-order statistical properties of interest except by empirical sampling, which cannot provide the insights that exact results can.

Here we derive an exact solution to the dead leaves model, by calculating joint probability distributions explicitly for arbitrary image features. This solution also provides a principled way to relate the features in a dead leaves image to the unobserved object attributes that cause these features. Since these relationships are precisely what we rely upon to see, this result thereby elevates the dead leaves model from an interesting approximation of natural images to a valuable tool for modeling perceptual inference and neural computation in the visual system.

To illustrate how this solution helps us understand natural scenes, we apply it to explain the highly non-gaussian probability distributions of two important types of image features: wavelet coefficients --- {\it i.e.} the image overlap with localized, oriented filters --- and local object boundaries. These features are important because they describe stimuli to which neurons in the early visual system are sensitive, and because high-order correlations between them reflect the physical objects and attributes in the visual world. Since the functional significance of neural responses to features can depend on the shape of the feature distribution \cite{Laughlin:1981p5727, Zetzsche:2005p6157, Sharpee:2007p5909}, it is important to understand why the distributions have their observed structure.

We first look specifically at the marginal, joint, and conditional distributions of wavelet coefficients. In natural images, the marginal distributions have heavy tails \cite{Mallat:1989p5575,Ruderman:1994p31,Wainwright:2000p4246}, which we show is due to the spatial scale invariance of objects. Joint and conditional distributions of wavelet coefficients have peculiar shapes (diamonds, pillows, bowties) that depend on the orientation and distance between the wavelets \cite{Simoncelli:1999p5457,Buccigrossi:1999p5408,Lee:2001p1}. We show how these distribution shapes arise naturally from occlusion by spatially extended objects. Finally, we compute the likelihood that a given pair of local object boundaries comes from the same physical contour. When estimated empirically from natural images, this likelihood predicts human judgments about contours \cite{Geisler:2001p808}. Our solution of the dead leaves model recovers the empirical statistics but only if one properly accounts for the relative depths at local boundaries, implicating depth cues in simple judgments about contours.

\section{Results}

\subsection{Solving the dead leaves model}
\label{exactsolution}

The pixels of a dead leaves image are fully determined by the properties of objects that are at least partially unoccluded. These properties are drawn independently from specified distributions over position, depth, size and shape, and texture. Texture can include both mean intensity and (possibly correlated) variations about the mean. When we say that we have solved the dead leaves model, we mean that we can calculate the joint probabilities of any model variables of interest, whether pixel intensities or object properties. This would be straightforward if the image components were related by linear superposition, but is much more difficult due to the strong nonlinearity of occlusion.

The essential property that makes the dead leaves model tractable is that different objects have independent attributes. Others have invoked the independence of object properties to derive the two-point correlation functions \cite{Ruderman:1997p825} and bivariate intensity probabilities \cite{Lee:2001p1} using a recursive argument that accounts for the way nearby objects occlude more distant ones. We were able to generalize this calculation from two points to an arbitrary collection of $N$ pixels, for which we can now calculate the multivariate joint intensity distribution. This distribution can then be transformed into feature probabilities, and related to the unobserved object properties.

If one samples the intensity of a particular dead leaves image at various locations, each pixel value will be determined by the texture of whichever object is at the top of the stack at that location. All pixels that fall into the same object share its texture, and are thereby correlated; pixels sampling from different objects are independent. Thus, if we can specify how the pixels are divided geometrically into objects, then we know the complete correlation structure for that image.

We can mathematically describe the configuration of objects at a given set of $N$ pixels by defining an {\em object membership function}, $\obj$, designating which pixels are `members' of which objects. (The symbol $\obj$ is the Hebrew letter {\it mem}, chosen to evoke the word {\it mem}bership.) In mathematical language, $\obj$ is a set partition of the $N$ pixels, so it is technically a set of sets: each set corresponds to a different object, and it contains the pixel locations at which that object is unobscured by any other objects. For example, one might find in a given image that pixels $\bx_1$, $\bx_2$ and $\bx_6$ fall into one object, $\bx_4$ and $\bx_5$ fall into a different object, and $\bx_3$ is alone in a third object (Figure \ref{exampleDeadLeavesImage}C). Then the corresponding object membership function can be expressed as $\obj=\{\{\bx_1,\bx_2,\bx_6\},\{\bx_3\},\{\bx_4,\bx_5\}\}$, or abbreviated as $\obj=\{126|3|45\}$.

The object membership function does not contain information directly about the intensities, but only about which pixels are correlated. Given a particular object membership $\obj$ for some selected pixels, the probability distribution $P(\bI|\obj)$ of image intensities $\bI$ factorizes into a product over objects: The different object textures are independent, and hence so are their respective pixels. In the above example, the probability distribution of intensities at those six pixels would be $P(\bI|\obj)=P(I_1,I_2,I_6|\obj)P(I_3|\obj)P(I_4,P_5|\obj)$. In general,
\begin{equation}
\label{factordistribution}
P(\bI|\obj)=\prod_{n=1}^{|\obj|}P(\bI_{\obj_n}|\obj)
\end{equation}
where $|\obj|$ is the number of objects, $\obj_n$ is the set of pixels falling into the $n$th object of $\obj$, and $\bI_{\obj_n}$ is a vector of intensities at those pixels. The factors $P(\bI_{\obj_n}|\obj)$ reflect the joint probabilities of intensities in a single, textured object. This formulation requires that we specify a texture model to provide these probabilities. For concreteness we use a simple gaussian white noise texture superposed on a uniform intensity (Methods), though any other probabilistic texture model could be used instead. Note that the texture model is wholely unrelated to the geometrical aspects of the dead leaves model.

If the geometric configuration of objects is not known, then the joint distribution of intensities $P(\bI)$ is an average over all possible configurations. The factorized conditional distributions of Equation \ref{factordistribution} are then combined in the weighted sum
\begin{equation}
\label{mixturedistribution}
P(\bI)=\sum_\obj P(\bI|\obj) P(\obj)
\end{equation}
This is a mixture distribution in which each mixture component $P(\bI|\obj)$ has a distinct correlation structure amongst pixels, induced by the different object membership functions. The weighting coefficients are {\it object membership probabilities} $P(\obj)$, i.e. the probability of observing the corresponding memberships over all possible dead leaves images with a given shape ensemble. Figure \ref{contourplots} shows examples of simple mixture distributions.

The object membership probability $P(\obj)$ represents how frequently the $N$ selected pixels are grouped into different objects according to $\obj$. We calculate each probability recursively, generalizing an argument of \cite{Ruderman:1997p825}. To do so, we must introduce some additional notation. We designate $\obj_{\setminus n}$ as the object membership function that remains after removing the $n$th object. We also define a boolean vector $\bbsigma(\obj,n)$ with $N$ components $\sigma_i(\obj,n)=(\bx_i\in\obj_n)$ that each indicate whether the pixel $\bx_i$ is contained in the $n$th object of $\obj$. For instance, $\bbsigma(\{126|3|45\},3)=(0,0,0,1,1,0)$.

By construction, there is a sequence of objects in any dead leaves image, ordered by depth. Consider only the topmost object. There is some probability, which we will denote by $Q_{\bbsigma(\obj,n)}$, that this top object includes all of the pixels in the set $\obj_n$, while excluding all the other pixels in $\obj_{\setminus n}$. Such an arrangement {\it partially} satisfies the membership constraint imposed by $\obj$. But for this object configuration to contribute to $P(\obj)$, we still need to ensure that the excluded pixels are also grouped appropriately by objects `deeper' in the image. The probability that deeper objects satisfy these reduced membership constraints is $P({\obj_{\setminus n}})$. Note that this probability is unaffected by whether the deeper objects would have enclosed the pixels in $\obj_n$: Objects at those positions are already occluded by the top object. There is also a probability $Q_{\bbsigma(\obj,0)}$ that the top object contains {\it none} of the $N$ selected pixels. Given this event, the probability of finding objects deeper in the stack that satisfy the membership constraints is just the original factor $P(\obj)$. Summing together all possibilities for the top object, we find $P(\obj)=Q_{\bbsigma(\obj,0)}P(\obj)+\sum_{n=1}^{|\obj|} Q_{\bbsigma(\obj,n)}P({\obj_{\setminus n}})$. Solving for $P(\obj)$ gives the recursion relation
\begin{equation}
P(\obj)=\frac{1}{1-Q_{\bbsigma(\obj,0)}}\sum_{n=1}^{|\obj|} Q_{\bbsigma(\obj,n)}P({\obj_{\setminus n}})
\label{recursion}
\end{equation}
Crucially, the image that remains below the top object is yet another dead leaves image, with all the same statistical properties as before, so we can calculate $P({\obj_{\setminus n}})$ by the same formula, recursively. Eventually the recursion terminates when there are no pixels left in $\obj$, with $P(\emptyset)=1$.

This recursive equation applies universally to any dead leaves model with independent, occluding objects, regardless of shape. In contrast, the factors $Q_{\bbsigma(\obj,n)}$ depend on the particular shape ensemble and the chosen set of pixels. In the Methods section we derive the general form of these factors for arbitrary shapes with smooth boundaries. The Supporting Information (Text \ref{bigcalc}) provides mathematical details of the calculation for the scale-invariant ensemble of elliptical shapes used from here onward.

The number of possible object membership functions quickly grows large as we consider more pixels. The limiting step is the number of possible object membership functions, known as Bell's number $B_N$, which unfortunately grows slightly faster than exponentially. In practice this restricts exact calculation to around a dozen pixels. Despite this limitation, interesting insights can be gained both by using few pixels or few subsets of possible object memberships, and by analyzing the general behavior in various limits. For instance, in low-clutter conditions when the maximal distance between pixels $u$ is much smaller than the minimum object size $r_-$ (e.g. Figure \ref{exampleDeadLeavesImage}A), object membership probabilities $P(\obj)$ behave as $P(\obj)\sim(u/r_-)^{|\obj|-1}$ (Figure \ref{Pmems}). Consequently, edges are rare ($|\obj_{\rm edge}|=2$), T-junctions are rarer ($|\obj_{\rm T-junc}|=3$), and every other feature is rarer still ($|\obj|>3$). In the sections below we use the solution of the dead leaves model and relevant approximations to explain complex statistical properties of natural scenes.

\subsection{Feature distributions}
\label{naturalscenestats}

In this section we calculate the joint feature probabilities in specific cases where features are linear functions of the pixel intensities, $f=F\bI$ for some filter matrix $F$. We set the filters of $F$ to be wavelets, local derivative operators (edge detectors) with a given orientation and scale. Choosing Haar wavelets, which weight intensities by $\pm 1$, emphasizes non-gaussianity of feature distributions and thereby establishes a more stringent test for the image model \cite{Lee:2001p1}.

It has been previously reported that empirical histograms of different Haar wavelets and wavelet pairs in the dead leaves model qualitatively reproduce the marginal and joint distributions in natural scenes \cite{Lee:2001p1}. Where empirical sampling can, at best, expose these interesting statistical similarities, our analytical results let us understand their origins.

\subsubsection{Marginal distributions of wavelet coefficients}

\begin{figure}[htb]
\centering
\includegraphics*[width=7in]{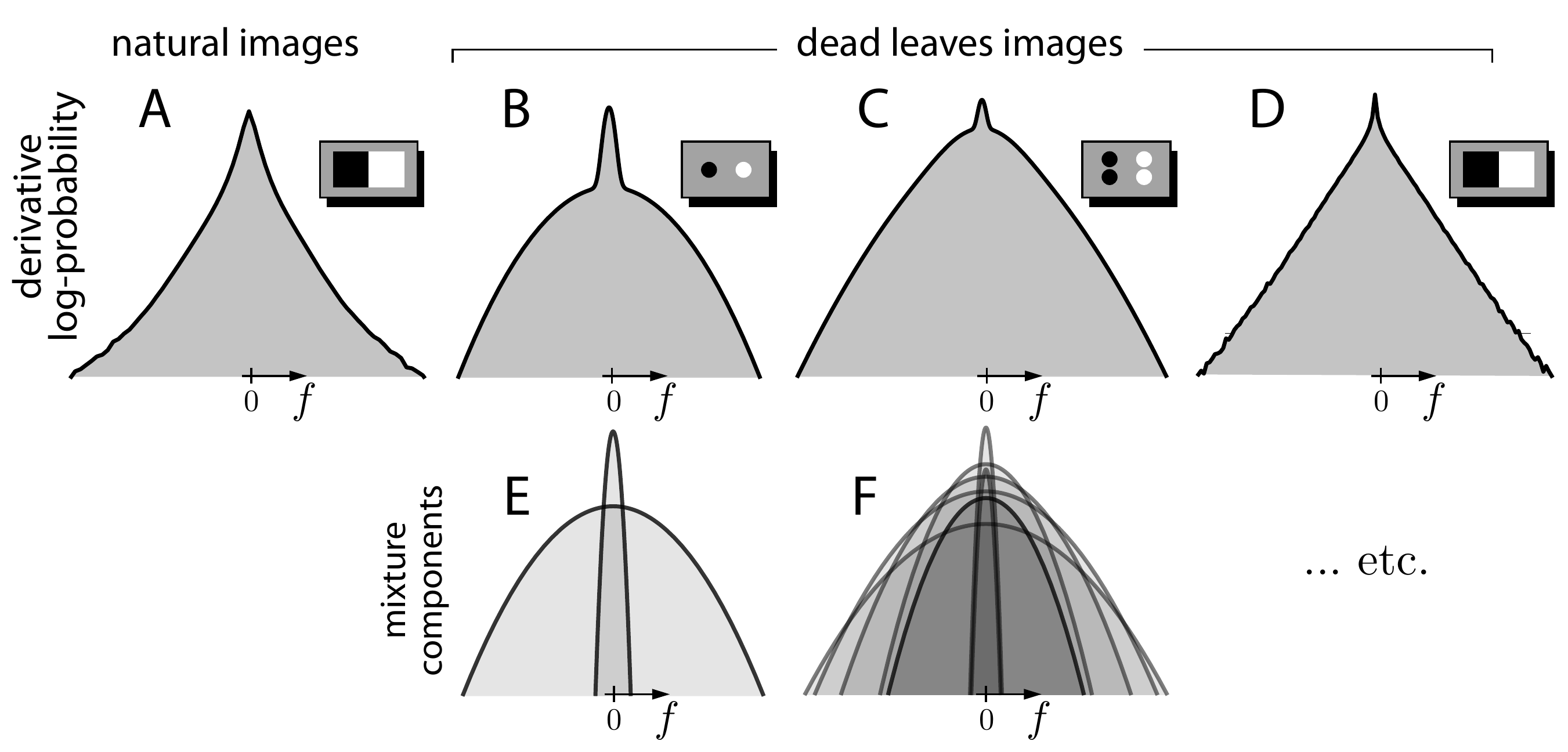}
\caption{Log-probability feature distributions $\log{P(f)}$ of spatial derivatives $f$. ({\bf A}) Empirically sampled distribution of derivatives (depicted graphically, inset) in natural images. ({\bf B}, {\bf C}) Feature probabilities calculated exactly for dead leaves images, using just one and two samples per image patch respectively (inset). ({\bf D}) Empirically sampled distributions for dead leaves images using a 16$\times$16 grid of samples per patch (inset). ({\bf E}, {\bf F}) Mixture components $P(f|\obj)$ corresponding to panels B and C respectively.}
\label{differences}
\end{figure}

One well-described feature of natural images is that the distribution of spatial derivatives $\Delta$ has heavy tails (Figure \ref{differences}A) well approximated as a generalized Laplace distribution $P(\Delta)\propto e^{-|\Delta|^\beta}$ for an exponent $\beta$ near 1 \cite{Mallat:1989p5575,Ruderman:1994p31,Wainwright:2000p4246,Lee:2001p1}. The heavy tails in these distributions cannot be obtained from a standard correlated gaussian model, because any projection of a multidimensional gaussian is again gaussian. Higher-order statistical structure is required.

This distribution can be calculated exactly for the dead leaves model by representing the local derivative by a simple feature: the intensity difference between nearby points, $f=I_1-I_2$. The resultant feature distribution is a mixture of two components, a narrow central peak and a broader tail (Figure \ref{differences}B,E). While this is a more kurtotic distribution than the gaussian texture, it does not closely match natural derivative histograms (Figure \ref{differences}A).

A simple consideration can account for the discrepancy. In our solution of the dead leaves model, what we have described so far as pixels are actually samples at infinitesimal points. In contrast, pixels in natural images represent light accumulated over some finite sensor area set by film grain, camera sensor wells, or photoreceptor cross-sections. This means that measured pixel values don't directly reflect an intensity sampled from an object but instead reflect integrals over unresolved sub-pixel details. When many samples are summed over some region $X_i$, one might na\"ively expect the total $\bar{I}_i=\sum_{j:\bx_j\in X_i} I_j$ to be gaussianly distributed. However, the usual central limit theorem does not apply, because of the correlations between the variables that is induced by spatially extended objects. These correlations can be segregated by re-expressing the total intensity in an image patch as a sum of the mean intensities in visible objects, weighted by their visible areas, $\bar{I}_i=\sum_{n=1}^{|\obj|}|\obj_n|J_n$ where $|\obj_n|$ is the visible area of the $n$th object and $J_n$ is the average intensity in that area. The summands $|\obj_n|J_n$ are approximately independent because they correspond to different objects. (They are not strictly independent because the visible areas $|\obj_n|$ are constrained to add up to the total area of the patch.) This way of writing $\bar{I}_i$ reveals two reasons why the sum does not converge to a normal distribution: the number of summands is a random variable, and the summands themselves have long-tailed distributions.

Scale invariance demands that the areas of homogeneous regions $|\obj_n|$ be distributed as a power law with exponent 2 \cite{Alvarez:1999p5954}. If the mean intensity within an object, $J_n$, is distributed more narrowly than this, then the distribution of visible areas $|\obj_n|$ will dominate the tail behavior of the products $|\obj_n|J_n$. $\bar{I}_i$ is thus a sum over a random number (one per visible object) of power-law distributed terms. A generalized central limit theorem holds that the distribution of such random sums is a two-sided exponential distribution when summands are power-law distributed with exponents of 2 or higher (\cite{Kotz:2001p4853}, Figure \ref{differences}D).\footnote{Technically, this theorem requires a geometric distribution for the number of summands. The observed distribution of the number of visible objects is not geometric, but it is similarly broad with a width of the same order as its mean, so a similar result should hold.} Indeed, Figure \ref{differences}D shows nearly straight tails on the log-probability plot. In natural images, areas of homogeneous regions are again distributed as power laws, but depending on the particular image the exponents can be slightly below 2 \cite{Alvarez:1999p5954}. Another generalized central limit theorem shows that under such conditions the distribution of the sum has slightly heavier tails \cite{Gnedenko:1972p3632}, as observed (Figure \ref{differences}A). While these considerations apply directly to the average pixel intensities, they pertain equally to intensity differences.

We can visualize how the heavy tails emerge by plotting feature distributions conditioned on various object configurations. When many different objects are visible, the independent object intensities tend to average out giving a narrow distribution; when few objects are visible, their areas are large, so the few object intensities are heavily weighted and their distribution is proportionately broad (Figure \ref{differences}F). Components with this spectrum of distribution widths all combine to give the mixture distribution heavy tails (Figure \ref{differences}B--D).

\subsubsection{Joint distributions of wavelet coefficients}

\begin{figure}
\centering
\includegraphics*[width=7in]{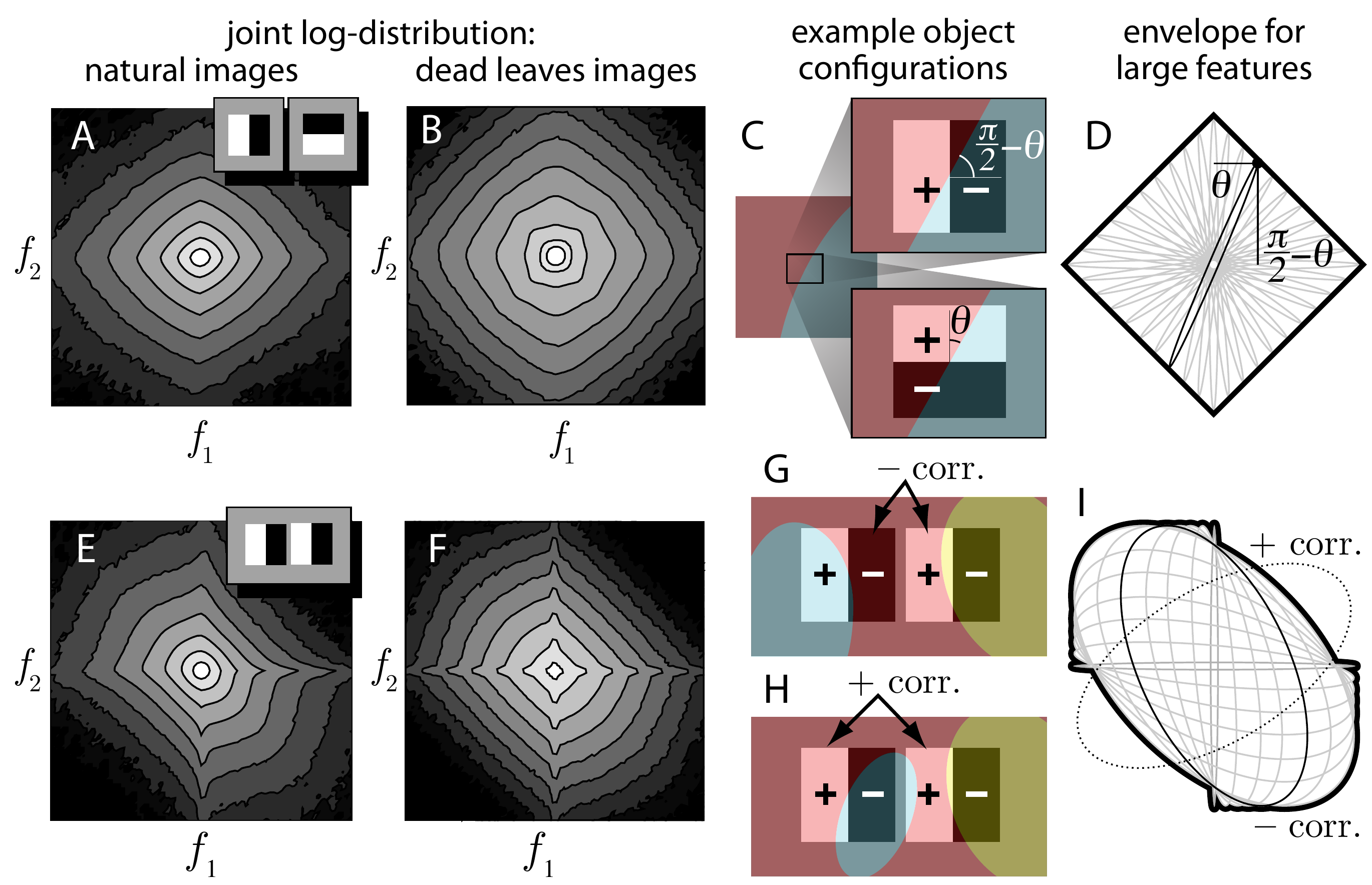}
\caption{Joint feature probabilities $\log{P(f_1,f_2)}$ for wavelet pairs $f_1$ and $f_2$. For orthogonal, colocalized Haar wavelets (inset of A, shifted for visibility), the contours of the empirically sampled bivariate distribution are diamond shaped for both natural images ({\bf A}) and dead leaves images ({\bf B}). At high feature amplitudes, certain object configurations have the greatest likelihood and thus dominate the joint distribution. Panel {\bf C} illustrates one such configuration. Colors indicate different objects with unspecified intensities. Dark and light shading show how the two wavelets weight the image pixels. ({\bf D}) Specifying only the object geometry (but not the object intensities) gives conditional feature distributions $P(f_1,f_2|\obj)$ that are bivariate gaussians with elliptical contours. For the conditional distributions that dominate at high feature amplitude, the contours trace a diamond-shaped envelope (thick curve) as a function of the relative angles between the object boundary and wavelet orientations (Text \ref{haarcovariance}).
Parallel, neighboring wavelets (inset of E) are anticorrelated, with joint probability contours exhibiting a similar `pillow' shape in natural images ({\bf E}) and dead leaves images ({\bf F}). Panels G and H illustrate object configurations that dominate at large feature amplitudes, colored as in C. ({\bf G}) If one object covers the opposite-sign lobes of neighboring wavelets while others prevent cancellation by the negative lobe, then the conditional feature distribution will have a negative correlation. ({\bf H}) Similarly, if one object covers the same-signed lobes of both features while another object prevents cancellation, then the conditional distribution will have a positive correlation. Configurations like this are much less probable than those like G, because the middle object must have precisely the right size and position. ({\bf I}) An ensemble of configurations like Panel G produce negatively correlated components (gray ellipses) that vary depending on how precisely the objects cover the feature lobes. The positively correlated components (dashed ellipse) caused by configurations like Panel H are many times less probable. Discounting the latter gives the mixture distribution an overall negative correlation, leaving components that trace out the pillow-shaped envelope (thick curve) seen in feature distribution contours (Methods).}
\label{jointwavelet}
\end{figure}

Within natural images, the feature distribution for two orthogonal, colocalized wavelets has diamond-shaped contours (Figure \ref{jointwavelet}A). Densely sampled dead leaves images reveal the same diamond contours (Figure \ref{jointwavelet}B); they are already visible when features are represented sparsely (Figure \ref{jointwaveletsupplement}A,B). For both natural images and dead leaves images, the distinctive non-gaussian structure is most visible for contours at large feature amplitude. In this limit, the mixture components with greatest likelihood dominate the distribution, and the most likely component at high amplitudes is the one with the greatest variance in the given feature direction. The greatest variance for a single Haar wavelet occurs when a boundary between two objects aligns with the boundary between oppositely-signed lobes, because that minimizes cancellation and maximizes the overlap each object makes with each lobe. However, this arrangement gives a minimal variance for the orthogonal wavelet at the same location. As the object boundary rotates (Figure \ref{jointwavelet}C), the overlap with one wavelet reduces by exactly the amount that the overlap increases for the orthogonal wavelet. Since a mixture component's width is proportional to the overlap, this perfect trade-off gives the maximum likelihood contours the observed diamond shape (Methods, Figure \ref{jointwavelet}D).\footnote{The slight squashing of the diamond shape seen for natural images is a consequence of gravity: in this natural image database there are more vertical contours than horizontal ones. Here the dead leaves model does not show this asymmetry because it is isotropic by construction.}

For neighboring Haar wavelets with the same orientation, there is once again a remarkable similarity between pillow-shaped distributions measured for natural images and dead leaves images (Figure \ref{jointwavelet}E,F). These too can be explained using simple arguments about the object geometry that dominates at high feature amplitudes. Negatively correlated mixture components occur when an object overlaps neighboring lobes on neighboring filters (Figure \ref{jointwavelet}G). Positively correlated components occur when an object covers the same-signed lobes of both Haar wavelets without cancellation by the intervening lobe of opposite sign. This can only happen if a small object occludes that oppositely signed lobe (Figure \ref{jointwavelet}H). Since a very limited variety of sizes and positions can accomodate this configuration, the positively correlated mixture components have much lower weights. Figure \ref{jointwavelet}I shows mixture components with negative correlations, from which emerge the basic `pillow' shape of the full bivariate feature distributions (Methods).

\subsubsection{Conditional distributions of wavelet coefficients}

\begin{figure}
\centering
\includegraphics*[width=7in]{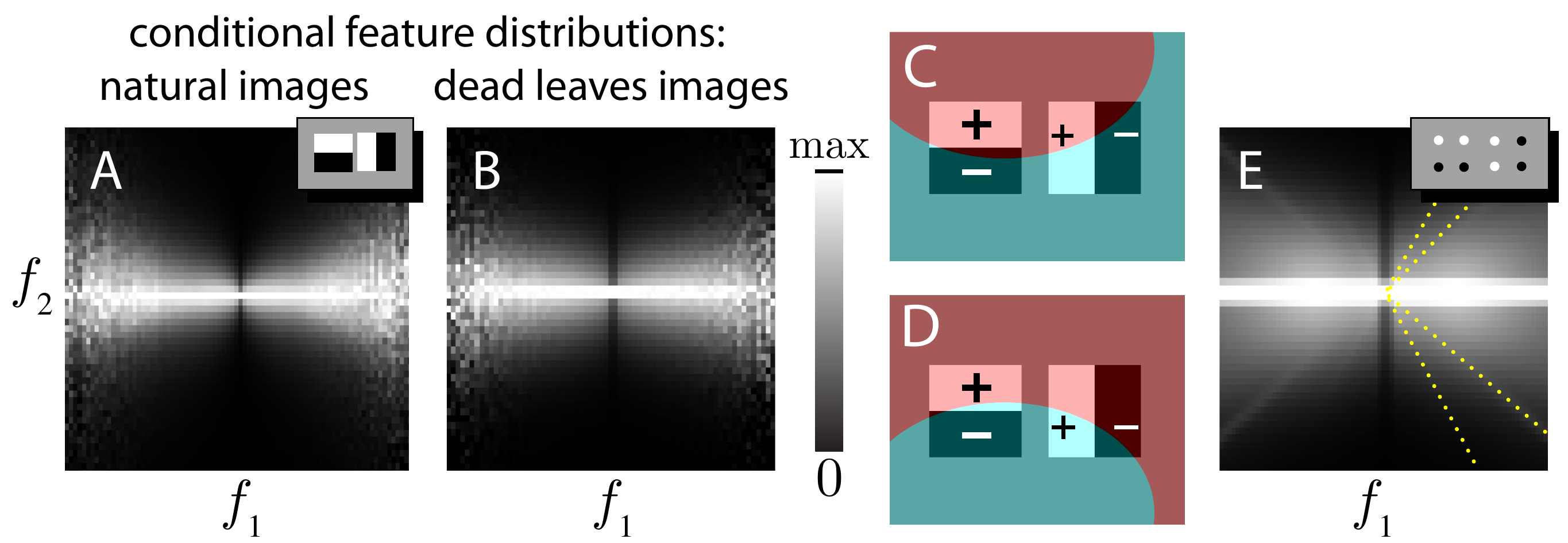}
\caption{`Bowtie' shapes appear in empirically sampled conditional feature distributions $P(f_2|f_1)$ for both natural images ({\bf A}) and dead leaves images ({\bf B}). Horizontal and vertical axes represent the coefficients of two neighboring, orthogonal Haar wavelet filters, $f_1$ and $f_2$ (inset of A). The grayscale is normalized so black represents 0 and white is the maximum probability for a given $f_1$. ({\bf C}, {\bf D}) Two equally probable object configurations, colored as in Figure \ref{jointwavelet}C, have identical $f_1$ but opposite $f_2$. Both features are proportional to the intensity difference between foreground and background objects. ({\bf E}) Conditional feature distribution with only four samples per feature (inset). Traces of the limited sampling appear as the faint diagonal bands passing through the origin (highlighted with dotted lines on right half). Each distinct band corresponds to a conditional distribution given a different object membership function, $P(f_2|f_1,\obj)$. Symmetry ensures that there will be no linear correlation between the two features, even as the width of $P(f_2|f_1)$ increases with $|f_1|$. With features sampled more densely, more such diagonal bands appear, until the bands blend together (B). This produces the distinctive bowtie shape in the conditional feature distributions.}
\label{bowties}
\end{figure}

Wavelet coefficients in natural scenes may be nearly decorrelated to second order yet still have a strong statistical dependency taking the form of a `bowtie'-shaped distribution of one filter coefficient conditioned upon another (Figure \ref{bowties}A) \cite{Simoncelli:1999p5457,Buccigrossi:1999p5408}. The dead leaves model reproduces this behavior (Figure \ref{bowties}B), and allows us to interpret it as well.

The distribution of intensities found within an object is narrower than the intensity distribution averaged over all objects. Consequently, when a wavelet filter lies across an object boundary, it typically yields a larger magnitude than the same filter applied wholely within a single object. Since object boundaries tend to extend across space, a second filter with different scale or orientation has an elevated probability of encountering the same edge. However, as Figs. \ref{bowties}C--D illustrate, the relative sign and magnitude of the two feature amplitudes depends on how the object boundary overlaps the second filter. In this symmetric example, positive and negative feature amplitudes are equally probable, so the conditional distribution $P(f_2|f_1)$ broadens with $|f_1|$ without any change in the mean (Figure \ref{bowties}E). This explains why the variability in one feature amplitude increases with the amplitude of a nearby feature.

\subsection{Shared causes of edges}
\label{sharedcauses}
A major advantage of using the dead leaves model is that the {\it causes} of image features --- objects and their attributes --- are represented explicitly. Our results relate these causes to each other as well as to the observable, pixel-based image features.

\begin{figure}
\centering
\includegraphics*[width=7in]{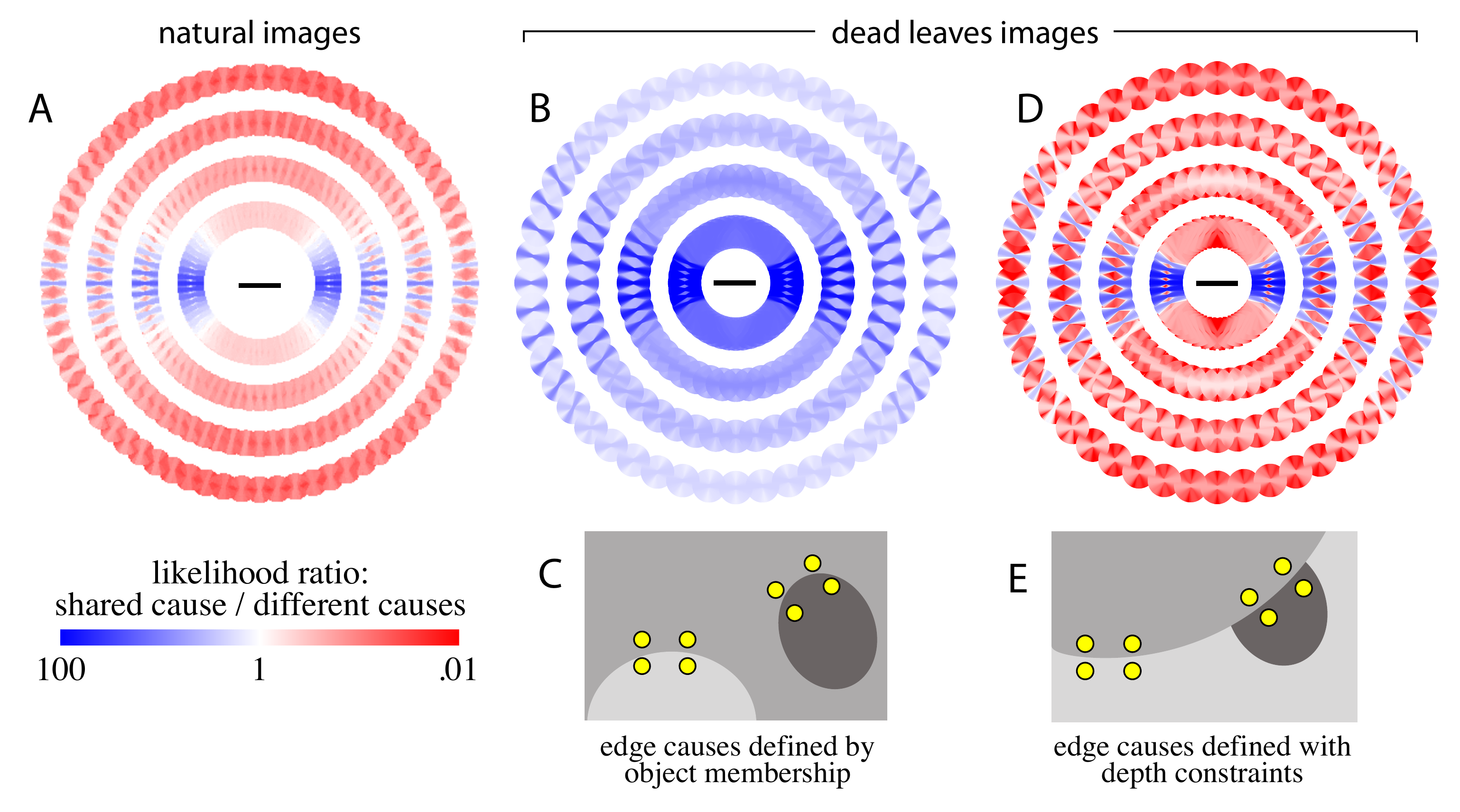}
\caption{Joint statistics of local edges and global contours. ({\bf A}) The likelihood ratio that edge pairs in natural images are caused by a common object versus by different objects (replotted from \cite{Geisler:2001p808} with permission). For test edges at many distances, directions, and orientations relative to a reference edge (horizontal bar at origin), line segments are colored to indicate the likelihood ratio (Methods). The segments are sorted so those indicating high likelihoods appear in front. Concentric white rings correspond to unsampled distances. In the dead leaves model, we can define the corresponding likelihood in one of two ways. First, a pair of edges could have a `shared cause' if at least one side of each edge samples from the same object. The resultant likelihood is shown in ({\bf B}) and an example of a shared cause is shown in ({\bf C}). Second, we may add a depth constraint to better describe the existence of a shared contour: this shared object must also be on top of the other objects. Using this second definition, panel ({\bf D}) shows the likelihoods and ({\bf E}) gives an example configuration. These likelihoods reproduce the observations made in natural images (A).}
\label{GeislerPlots}
\end{figure}

In natural images, edge pairs tend to fall tangent to circles passing through both edge locations \cite{Sigman:2001p846,Chow:2002p1048}. Geisler {\it et al.}\ \cite{Geisler:2001p808,Geisler:2009p5375} augmented such an analysis with global information about physical contours, by laboriously hand-segmenting objects within many images of foliage. The likelihood that two edges share a physical cause (Figure \ref{GeislerPlots}A) --- i.e. belong to the same contour --- were highly predictive of human judgements of whether the edges had a shared cause.

The dead leaves model can provide a mathematical `ground truth' for such calculations. First, we represent individual edge features by an object membership function that divides four pixels into two pairs (Figure \ref{geislerplotSI}A). Second, we define the conditions under which a pair of edges have the same physical cause. Third, for edge pairs with various geometrical relationships (Figure \ref{geislerplotSI}B) we plot the likelihood ratio under the hypotheses of a shared cause versus different causes (Methods).

A seemingly natural condition would identify a shared cause when there exists an object that participates in both edges. The resultant likelihood ratio always favors a shared cause, for all relative positions and orientations of the edge pair (Figure \ref{GeislerPlots}B), at odds with reported statistics (Figure \ref{GeislerPlots}A) \cite{Geisler:2001p808,Geisler:2009p5375}. The reason can be seen in Figure \ref{GeislerPlots}C: An object could be shared across two edges simply if it is a common background for two distinct objects. Thus this definition, only involving object identity on both sides of an edge, is inadequate to reproduce the observed edge statistics.

A more sensible pattern emerges by modifying the definition of common cause to include relative depth, assigning `border ownership' \cite{Zhou:2000p1089} to the local edge. We now define a common cause to exist when a single object participates in both edges, {\it and} is closer to the viewer than the other objects seen at these edges. An example of this configuration is seen in Figure \ref{GeislerPlots}E, which agrees with our intuition about a shared cause for two edges. Application of this definition requires that the object membership function be augmented to include the objects' relative depths, yielding an {\it ordered} object membership function. Their probabilities can be calculated by a very similar recursion equation as that used for the unordered variant (Methods). With this definition, Figure \ref{GeislerPlots}D shows that certain edges are more likely to have a common cause, whereas other edges are more likely to be independent. The pattern closely resembles results of Geisler {\it et al.}\ \cite{Geisler:2001p808,Geisler:2009p5375} (Figure \ref{GeislerPlots}A). Since those statistics were predictive of human judgments about contour completion across occluders, therefore the dead leaves model also qualitatively predicts human inference about such ambiguous stimuli.

\section{Discussion}

Our study used an occlusion model to explain several distinctive statistical regularities in natural images. The model describes images composed of many independent, opaque objects. We solved this image model by deriving exact probability distributions that relate arbitrary image features to each other and to the depicted objects. By applying and analyzing this solution we were able to account for several curious observations about image features, summarized very briefly as follows. We saw that heavy-tailed feature distributions are explained by integrating over sub-pixel details with scale-invariant spatial structure (Figure \ref{differences}). The diamond-shaped joint distribution of orthogonal, colocalized wavelets occurs because edges aligned well with one wavelet must be aligned poorly with an orthogonal wavelet (Figure \ref{jointwavelet}). The pillow-shaped joint distribution of parallel wavelets reflects the rarity with which objects can induce positive correlation by squeezing precisely into one wavelet lobe (Figure \ref{jointwavelet}). Bowtie-shaped conditional distributions arise because extended object boundaries can overlap wavelets with identical amplitudes but opposite signs (Figure \ref{bowties}). Finally, accurately computing the likelihood that two edges share a physical cause depends critically on ascribing relative depth to the edges (Figure \ref{GeislerPlots}). The unifying idea is that seemingly complex statistics of edge features can be explained by simple geometric configurations of a few opaque objects.

These results were made possible by connecting image features to object configurations through the object membership function $\obj$. This representation enables probability distributions to be decomposed into a mixture of simpler distributions. The existence of a mixture distribution for the dead leaves model was first proved in \cite{Bordenave:2006p2952,Gousseau:2007p2970}. Here we found an explicit solution for the mixture components that yields concrete numbers used in the applications above. Additionally, this solution generalizes to give probabilistic relationships among all model variables (Section \ref{generalizations}), including object texture, size, shape, position, and depth. The ability to relate arbitrary image features and many diverse object attributes in a principled manner is a substantial advance over previous efforts.

Although occlusion is a ubiquitous and fundamental attribute of natural scenes, it is not the sole process that could cause these effects. However, our results should generalize to other processes that share crucial attributes: only one physical cause dominates the image at each point, and separate causes are drawn from a scale-invariant size distribution. As one striking example, the cratered lunar surface appears remarkably similar to dead leaves images \cite{StuartAlexander:1978p5726}. Even though the causal process is entirely different from occlusion, the essential properties are identical: New impacts locally erase traces of previous impacts, and small craters are much more common than large ones. Similar principles may approximate other physical processes as well, such as those that determine surface composition or some three-dimensional bump textures. The results presented here should pertain to feature statistics caused by any such `exclusion' process.

\subsection{Beyond the dead leaves model}

Despite the dead leaves model's success at reproducing many complex natural statistics, we expect some statistical differences also. Indeed, whereas natural scenes appear reasonably gaussian after normalizing intensities by the local standard deviation \cite{Ruderman:1994p31,Wainwright:2000p4246}, dead leaves images do not have this property. This therefore excludes object boundaries as the cause of this property, despite speculations to the contrary \cite{vanHateren:1997p6084}. By extending the model in various ways, one may hope to capture this and other natural image properties and thereby reveal their underlying cause.

Most real objects have more elaborate shapes than the ellipses used in these calculations. Notably, the most common edge configuration seen in natural scenes is consistent with circular \cite{Sigman:2001p846}, elliptical or parabolic \cite{Geisler:2009p5375} arcs. This accounts for why the elliptical object ensemble could reproduce statistics of images populated by complex, natural objects. Incorporating more complex objects may correct some minor discrepancies between the dead leaves model and natural scenes.

The realism of the dead leaves model could be further improved by adding correlations between model variables. For instance, light sources could be modeled by modulating texture according to position within each object. Rudimentary three-dimensional shape could be included using textures to indicate object tilt \cite{Saunders:2001p4794}. Perspective could be modeled by covarying size with depth. Binocular disparity could be included by generating image pairs in which every object has a positional shift coupled to its depth. Images with such improvements could be easily generated, but in some cases a new solution for the enhanced model would be required.

\subsection{Toward neural coding of natural scenes}

Some perceptual tasks can be accurately modeled as inference based on simple models of stimulus probabilities \cite{Ernst:2002p848,Battaglia:2003p833,Kording:2004p842,Howe:2005p5485,Stocker:2006p809}. Human perception of images appears biased toward statistically probable features of the dead leaves model. For example, empirical edge statistics predict psychophysical judgments about whether two edges have a common cause \cite{Geisler:2001p808}, and the dead leaves model reproduces these statistics. Artificial neural networks trained on dead leaves images make systematic interpretation errors that are consistent with illusory percepts in humans \cite{Corney:2007p4462}. Such evidence hints that these percepts might result from perceptual inference using probabilities described by the dead leaves model.

On a more mechanistic level, some electrophysiological recordings of individual neurons in animal cortex appear consistent with a probabilistic weighing of sense data \cite{Murray:2002p834,Mazurek:2003p829,Ma:2006p839}. We might speculate that some cortical neurons could be tuned to encode feature probabilities. For instance, V1 complex cells are excited by edges irrespective of polarity and precise location of those edges \cite{Hubel:1962p821}, and are especially sensitive to phase alignment caused frequently by object boundaries in natural images \cite{Felsen:2005p5896}. We might therefore wish to describe a rudimentary complex cell as encoding the probability that an edge passes through two points in its receptive field, irrespective of which side of the edge is brighter. In our formalism, this corresponds to an object membership function $\obj_{\rm edge}=\{1|2\}$.  Assuming that objects have gaussian-distribution intensities and the image sensors have some additive gaussian noise, the probability of an edge given the intensity difference $\Delta$ across space is $P(\obj_{\rm edge}|\Delta)=\left[1+k\exp{\left(-\beta\Delta^2\right)}\right]^{-1}$, where $k$ and $\beta$ are positive constants that depend on the spatial scale, overall image contrast, and sensor noise. This function resembles the contrast-energy model of complex cells \cite{Adelson:1985p831} with a saturating nonlinearity. Thus we might interpret complex cell activity as encoding the probability of a local edge in a world of objects. It will be interesting to explore such a model more thoroughly, and to see if other neurons have properties that map nicely onto representations of still more complex features within the dead leaves model. Since synaptic connections are modified by neural correlations, and the occlusion model explains stimulus correlations, therefore the model may also help generate predictions about cortical circuitry that has matured in the natural world.

In vision science, progress has been made by finding stimuli appropriate for the area of study \cite{Rust:2005p1822}. The best stimulus is one that contains a rich repertory of the right kinds of features, while limiting extraneous detail. Since the dead leaves model shares many low- and mid-complexity features with the natural environment while simplifying some higher-level features, it seems like an especially good stimulus to use in experiments that probe the mechanisms of low- and mid-level vision. It strikes a good balance between tractability, accuracy, and richness, by isolating two causes of image features which must be disambiguated to interpret truly natural scenes: occlusion and texture. The availability of an exact solution for the relevant probabilities is a promising new ingredient for experimental and theoretical studies of visual function.

%

\section{Methods}

\small

\subsection{Dead leaves membership probabilities}
\label{inclusionprobs}

Equation \ref{recursion} expresses the object membership probabilities $P(\obj)$ in terms of some geometric factors $Q_{\bbsigma(\obj,n)}$. These factors represent the probability that points $\bx_i\in\obj_n$ are included in one object while the other points $\bx_i\in\obj_{\setminus n}$ are not, averaged over all object positions and shapes. For convenience, we name these $Q_\bbsigma$ `inclusion probabilities'. Note that these quantities involve the geometry of single objects only; the recursion of Equation \ref{recursion} converts them into the multi-object probabilities $P_\obj$ that characterize the dead leaves model geometry. In this section we show how the inclusion probabilities can be calculated for arbitrary objects.

We begin by specifying a shape through a `leaf' function $L_\sigma(\bx,\rho)$, which is an indicator function over space $\bx$ and shape parameter(s) $\rho$. The function can indicate either the inside or the outside of an object centered on the origin, depending on the binary variable $\sigma\in\{0,1\}$: $L_\sigma(\bx,\rho)$ equals $\sigma$ when pixel $\bx$ is inside the object and $1-\sigma$ when $\bx$ is outside it (Figure \ref{QsFigure}A). With this definition,
\begin{equation*}
Q_{\bbsigma(\obj,n)}(\bc,\rho)=\prod_{i=1}^N L_{\sigma_{i}(\obj,n)}(\bx_i-\bc,\rho)
\end{equation*}
is the inclusion probability that a leaf with shape $\rho$ and location $\bc$ includes all sample points $\bx_i\in\obj_n$ and excludes all remaining $\bx_i\in\obj_{\setminus n}$ (Figure \ref{QsFigure}B).

The inclusion probabilities $Q_{\bbsigma}$ in Equation \ref{recursion} are averages over all possible object shapes and positions. Thus we are interested in the average of Equation $Q_{\bbsigma}(\bc,\rho)$ over the distribution of leaf positions $P(\bc)$ and shapes $P(\rho)$:
\begin{equation*}
Q_{\bbsigma}=\int d\rho\,P(\rho)Q_{\bbsigma}(\rho)=\iint d\rho\, d\bc\,P(\rho)P(\bc)Q_{\bbsigma}(\bc,\rho)
\end{equation*}
We first perform the average over object positions $\bc$ to obtain $Q_{\bbsigma}(\rho)$, and subsequently calculate the average over object shape $\rho$.

In the dead leaves model, objects are distributed with uniform probability across space. For simplicity we also assume wraparound boundary conditions and with no loss of generality require that no object is larger than the image to avoid self-intersections. (We can allow larger objects by choosing a small window into the dead leaves world to represent our image; objects may be larger than the window but smaller than the entire model world.) By scaling distance so the image has unit area, we have $P(\bc)=1$ and the $\bc$-integral of binary-valued $Q_{\bbsigma}(\bc,\rho)$ gives the inclusion probabilities for a given $\rho$ as the areas of the regions with constant $Q_{\bbsigma}(\bc,\rho)$.

Direct integration is not straightforward even for simple object shapes because these regions generally have complicated two-dimensional limits and may not even be simply connected. However, using the divergence theorem we can transform this area integral into a simpler contour integral that follows object boundaries piecewise. The vector field ${\bf V}=\frac{1}{2}\bc$ has divergence (in $\bc$-space) of $\nabla\cdot{\bf V}=1$, so integrating this divergence over the desired region gives the enclosed area. The divergence theorem says that this integral equals the flux of $\bf V$ across the region boundary:
\begin{equation}
Q_{\bbsigma}(\rho)=\int d\bc\,P(\bc)Q_{\bbsigma}(\bc,\rho)=\int\limits_{C}\nabla\cdot {\bf V}\, d\bc=\oint\limits_{\partial C} {\bf V}\cdot \hat{\bf n}\, ds
\label{divergencetheorem}
\end{equation}
where $C$ is the region in $\bc$-space where $Q_{\bbsigma}(\bc,\rho)=1$, $\partial C$ is its boundary, $\hat{\bf n}$ is the unit normal vector to the boundary, and $ds$ is the arclength. The boundary is composed of piecewise smooth segments of the object outline centered on the sample points $\bx_i$ (Figure \ref{QsFigure}B). We index the relevant segments by $m\in M$, and represent the curves by $\bs_m(t):t'_m<t<t''_m$ for $t$ between the cusps at which the contour changes direction abruptly. The integral along each segment is then
\begin{equation}
\label{contourintegral}
A_m=\frac{1}{2}\int_{t'_m}^{t''_m} \bs_m(t)\cdot\hat{\bn}_m(t)\,ds
\end{equation}
and the complete contour integral is a sum over segments $Q_{\bbsigma}(\rho)=\sum_{m\in M}A_m$.

To average $Q_{\bbsigma}(\rho)$ over the shape ensemble $P(\rho)$ we need to compute $\int Q_{\bbsigma}(\rho)P(\rho)d\rho$. Note that the set of piecewise smooth segments composing the contour $\partial C$ may change depending on $\rho$, so the $\rho$-integral must itself be done piecewise. We define an index $\ell$ specifying the regions $R_\ell$ in $\rho$-space where a given set of segments $M_\ell$ compose the contour. Within $R_\ell$ the integral over $\rho$ can then be carried out on each summand $A_m$ separately, yielding
\begin{equation*}
Q_\bbsigma=\sum_{\ell}\sum_{m\in M_{\ell}}\ \int\limits_{R_\ell} d\rho\ P(\rho)A_m(\rho)
\end{equation*}

Carrying out this calculation explicitly, not just formally, requires some careful geometry. In the Supporting Information we complete these calculations for an ensemble of elliptical objects with an inverse-cube power-law distribution of sizes (Text \ref{bigcalc}). In principle it is also possible to calculate all these probabilities exactly for various other shape ensembles with simple boundaries such as polygons, or compound objects comprising multiple circles. Other size ensembles can also be used. The mathematical techniques required to complete the calculations are essentially the same.

For the figures presented in this paper, all objects were ellipses with uniformly distributed eccentricities between 1 and 4, uniformly distributed orientations, and an inverse-cube size distribution with upper and lower bounds $r_+=100$ and $r_-=1$. For Figs. \ref{differences}--\ref{bowties}, we used high-clutter conditions by setting the pixel spacing to $5r_-$. For Figure \ref{GeislerPlots}, to replicate the relatively low-clutter conditions under which the natural image statistics were measured empirically \cite{Geisler:2001p808}, we chose the pixel spacing to be $r_-/5$.

\subsection{Intensity and feature distributions}
\label{texturemodel}
For simplicity we assume that every object has a constant gaussian-distributed mean intensity and an additive gaussian white noise textural modulation with variances $\Xi_0$ and $\Xi_1$. For this texture ensemble, the conditional distribution of pixel intensities is $P(\bI|\obj)\propto\exp{\left(-\frac{1}{2}\bI^\top C_\obj^{-1}\bI\right)}$, with zero mean and covariance $\left(C_\obj\right)_{ij}=\Xi_0\sum_{n=1}^{|\obj|}\sigma_i(\obj,n)\sigma_j(\obj,n)+\Xi_1\delta_{ij}$. In the results shown in this paper, $\Xi_0=1$ and $\Xi_1=0.01$.

For features specified as linear combinations of intensities by ${\bf f}=F\bI$, the conditional distribution is $P({\bf f}|\obj)\propto\exp{\left(-\frac{1}{2}{\bf f}^\top(FC_\obj F^\top)^{-1}{\bf f}\right)}$ and the joint probability is the mixture distribution $P({\bf f})=\sum_\obj P(\obj)P({\bf f}|\obj)$.

\subsection{Averaging over image patches}
\label{haarcovariance}

Pixels in natural images are integrals of light intensity over a finite solid angle. In the dead leaves model, we can approximate these spatial integrals by summing over multiple points within an image patch $X_i$, defining
\begin{equation*}
\bar{I}_i=\sum_{j:\bx_j\in X_i} I_{j}
\end{equation*}
Using the white-noise texture model (Methods \ref{texturemodel}), the total intensity $\bar{I}_i$ over an image patch has a conditional distribution $P(\bar{I}_i|\obj)$ which is gaussian with zero mean and variance
\begin{equation*}
\sigma^2_{\bar{I}_i|\obj}=\sum_{jk}(C_\obj)_{jk}=\sum_{n=1}^{|\obj|}|\obj_n|^2\Xi_0+N\Xi_1
\end{equation*}
Here $C_\obj$ is the covariance matrix of all pixels in image patch $X_i$ conditioned on the object membership function $\obj$, and $|\obj_n|$ is the number of sampled pixels falling into the $n$th object. Thus the variance increases with the square of the sampled area of each object, and is maximized when only one object covers the sampling area.

A Haar wavelet takes the difference $H=\bar{I}_{1}-\bar{I}_{2}$ between sums $\bar{I}_{1}$ and $\bar{I}_{2}$ over two distinct regions (Figure \ref{overlapapprox}A). The corresponding variance does not necessarily increase with the square of each object's sampled area, because some of the samples are weighted with opposite signs and thus cancel. The conditional covariance between two Haar wavelets $H_i$ and $H_j$ is
\begin{equation}
C_{H_iH_j|\obj}=\sum_{n=1}^{|\obj|}\left(|\obj_n^{1,i}|-|\obj_n^{2,i}|\right)\left(|\obj_n^{1,j}|-|\obj_n^{2,j}|\right)\Xi_0+N_{ij}\Xi_1
\label{haaraveragecov}
\end{equation}
where $|\obj_n^{k,i}|$ is the number of samples in region $k$ of wavelet $i$ which fall into the $n$th object (Figure \ref{overlapapprox}A), and $N_{ij}$ is the number of samples shared by wavelets $H_i$ and $H_j$.

In Figure \ref{jointwavelet}B,D, the diamond-shaped contours emerge as a consequence of Equation \ref{haaraveragecov}. Instead of the Haar wavelets with square support shown in that figure, it is simpler to understand the case with circular support (Figure \ref{overlapapprox}B), though the result is the same. The maximum amplitude features occur when a single object boundary passes through the center of the wavelet at an angle $\theta$. The covariance of the mixture distribution conditioned on this object configuration is
\begin{equation*}
C_{HH|\theta}= 2N^2\Xi_0\left(\begin{array}{cc}
(\pi-2\theta)^2 & 2\theta(\pi-2\theta) \\
2\theta(\pi-2\theta) & (2\theta)^2 \\
\end{array}\right)+N\Xi_1{\bf 1}
\end{equation*}
where $N$ is the number of samples in each Haar wavelet. For large $N$ this covariance matrix is nearly singular, with almost unity correlation coefficient between the variations along $H_1$ and $H_2$. Contours of the corresponding bivariate gaussian have maximum extent at feature amplitudes proportional to $(\pm\theta,\pm(\frac{\pi}{2}-\theta))$. The envelope of these contours produces the diamond shown in Figure \ref{jointwavelet}B,D.

In Figure \ref{jointwavelet}, two neighboring, parallel Haar wavelets have a joint distribution with a distinctive `pillow' shape. The dominant contributions at high feature amplitudes involve three objects as depicted in Figure \ref{jointwavelet}G, one covering the left edge of the wavelet, a second one covering the right edge, and a third covering the gap between them. We can approximate this arrangement with a one-dimensional version, considering only the horizontal extent of objects (Figure \ref{overlapapprox}C,D). If we denote how much the leftmost and rightmost objects overlap the wavelets by $d_l$ and $d_r$, then the covariance of the mixture distribution is
\begin{equation*}
C_{HH|d_l,d_r}= N^2\Xi_0\left(\begin{array}{cc}
2\Delta_l^2 & \pm\Delta_l\Delta_r \\
\pm\Delta_l\Delta_r & 2\Delta_r^2 \\
\end{array}\right)+N\Xi_1{\bf 1}
\end{equation*}
where $\Delta_i=\min(d_i,1-d_i)$ and the width of each lobe of the Haar wavelet is $1$. These components all have a correlation coefficient of nearly $\pm 1/2$ but have different variances. By changing $d_l$ and $d_r$ for the configuration shown in Figure \ref{overlapapprox}C we obtain conditional distributions with the ensemble of contours seen in Figure \ref{jointwavelet}I. Their envelope produces the `pillow' shape (Figure \ref{jointwavelet}).

\subsection{Shared causes of edges}
\label{edgeMethods}

To define oriented edges, we select four pixels arranged in a rectangle, and select only those object membership functions that bisect these four pixels into two pairs. Note that a range of object boundaries can produce such a separation. Giving the rectangle an aspect ratio 2.75 constrains edges to an allowed range of orientations $2\tan^{-1}{(1/2.75)}=40^\circ$ (Figure \ref{geislerplotSI}A) that matches the orientation bandwidth of used in \cite{Geisler:2001p808}. Pairs of edges are described by two such bisected four-pixel clusters (Figure \ref{geislerplotSI}B). This definition of edge pairs restricts these eight pixels to have one of only seven possible object membership functions (Table \ref{commoncauses}A). In one of these configurations, every pixel pair is a member of a different object: $\obj=\{12|34|56|78\}$. In the remaining configurations, at least two pairs are members of the same object (Figure \ref{GeislerPlots}D). This latter category serves as one possible definition of a `shared cause' for the two edges.

A second definition of shared cause invokes not just the object membership but also the relative depth of the objects. In particular, we use {\it ordered} membership functions $\oobj$ (Section \ref{generalizations}), and we classify these $\oobj$ according to whether a pair of pixels from each edge both falls into the same object {\it and} that object is above the object present at the remaining pixels (Figure \ref{GeislerPlots}E). The relevant $\oobj$ are listed in Table \ref{commoncauses}B.

With either definition, the likelihood ratio of shared cause to different cause is $L={\sum_{\obj\in S}P(\obj)}/{\sum_{\obj\in D}P(\obj)}$, where $S$ and $D$ are the sets of membership functions categorized as shared or different causes respectively. This likelihood ratio varies as a function of the positions and relative orientation of the two edge pairs (Figs. \ref{GeislerPlots}C--D).

\begin{table}[htbp]
\centering
\begin{tabular}{| r | l |}
\multicolumn{2}{c}{{\bf A}: Classification of unordered $\obj$}\\
\hline
S: Shared cause & D: Different causes\\
\hline
1256 $|$ 34 $|$ 78 &12 $|$ 34 $|$ 56 $|$ 78 \\
1278 $|$ 34 $|$ 56 &\\
12 $|$ 56 $|$ 3478 &\\
12 $|$ 78 $|$ 3456 &\\
1256 $|$ 3478 &\\
1278 $|$ 3456 &\\
\hline
\end{tabular}

\begin{tabular}{| r | r c l |}
\multicolumn{4}{c}{}\\
\multicolumn{4}{c}{{\bf B}: Classification of ordered $\oobj$}\\
\hline
S: Shared cause & \multicolumn{3}{c|}{D: Different causes}\\
\hline
1256 $>$ 34 $|$ 78 & 34 $|$ 78 $>$ 1256 & 34 $>$ 1256 $>$ 78 & 78 $>$ 1256 $>$ 34 \\
3478 $>$ 12 $|$ 56 & 12 $|$ 56 $>$ 3478 & 12 $>$ 3478 $>$ 56 & 56 $>$ 3478 $>$ 12 \\
1278 $>$ 34 $|$ 56 & 34 $|$ 56 $>$ 1278 & 34 $>$ 1278 $>$ 56 & 56 $>$ 1278 $>$ 34 \\
3456 $>$ 12 $|$ 78 & 12 $|$ 78 $>$ 3456 & 12 $>$ 3456 $>$ 78 & 78 $>$ 3456 $>$ 12\\
1256 $|$ 3478 && 12 $|$ 34 $|$ 56 $|$ 78 &\\
1278 $|$ 3456 &&&\\
\hline
\end{tabular}
\caption{Object membership functions used for joint edge statistics. For compactness we represent object membership functions by the pixel indices divided symbolically into ordered or unordered groups. For example, $\{\{\bx_1,\bx_2\},\{\bx_3,\bx_4\}\}$ is written as $12|34$ if unordered, and as $12>34$ if ordered such that the object containing points $\bx_1$ and $\bx_2$ lies above the object containing $\bx_3$ and $\bx_4$. These object membership functions are classified according to whether they reflect a shared cause or different causes for the two edges, using unordered ({\bf A}) or ordered ({\bf B}) representations.}
\label{commoncauses}
\end{table}

\subsection{Generalizations}
\label{generalizations}
We can calculate the relative depth of objects by using an {\it ordered} object membership function $\oobj$ rather than an unordered membership function $\obj$. (The Hebrew letter {\it final mem} $\oobj$ is used only at the end of a word, representing that object order matters.) Whereas $\obj$ was a set of subsets, $\oobj$ is an ordered set of subsets with $\oobj_n$ representing the pixels contained by the $n$th-highest object sampled by any of the $N$ selected pixels. The recursion in this case is even simpler than Equation \ref{recursion}:
\begin{equation*}
P(\oobj)=\frac{1}{1-Q_{\sigma(\oobj,0)}}Q_{\sigma(\oobj,1)}P(\oobj_{\setminus 1})
\end{equation*}
There is no summation here because there is only one term for which the first object is highest in the stack of objects. One may use a partial ordering if not all relative depths are of interest, and then there will be a sum over arrangements consistent with the partial ordering.

Note that there are more hidden variables of interest besides the object membership and relative depth, and the joint probabilities of these can be calculated by a similar recursive formula, without marginalizing away the hidden variables. The joint distribution of shape and membership, for instance, can be calculated as
$$
P(\obj,\rho)=\frac{1}{1-Q_{\bbsigma(\obj,0)}}\sum_{n=1}^{|\obj|}P(\rho_n)Q_{\bbsigma(\obj,n)}(\rho_n)P(\obj_{\setminus n},\rho_{\setminus n})
$$
where $\rho$ is now a vector of $N$ shape parameters, with $\rho_n$ indicating the shape parameters for the topmost object present at pixel location $\bx_n$.

\subsection{Empirical sampling of dead leaves and natural images}
For probabilities involving many image points, we generate many dead leaves images and empirically sample from them to obtain histograms. Images are produced by layering objects from front to back until all image pixels are members of some object, a process that yields stationary image statistics \cite{Kendall:1999p5650}.

Natural images were drawn from van Hateren's image database \cite{vanHateren:1998p5554}. Feature distributions were obtained by log-transforming images \cite{Lee:2001p1}, filtering them by the relevant Haar wavelets, and computing univariate or bivariate histograms.


\section*{Acknowledgments}
The author thanks Ken Miller, Larry Abbott, Stefano Fusi, Taro Toyoizumi, Vladimir Itskov, and Tony Movshon for helpful comments and suggestions. This work was supported by the National Institute of Health Grant EY13933 and the Swartz Foundation.

\bibliographystyle{plos}
\bibliography{DeadLeavesBib5}

%
%

\renewcommand \thesection{S\arabic{section}}
\renewcommand \thefigure{S\arabic{figure}}
\renewcommand \thetable{S\arabic{table}}
\renewcommand \theequation{S\arabic{equation}}
\setcounter{figure}{0}
\setcounter{table}{0}
\setcounter{section}{0}
\setcounter{equation}{0}


\section*{}
\label{supportingfigures}

\begin{figure}[p!]
\centering
\includegraphics*[width=5in]{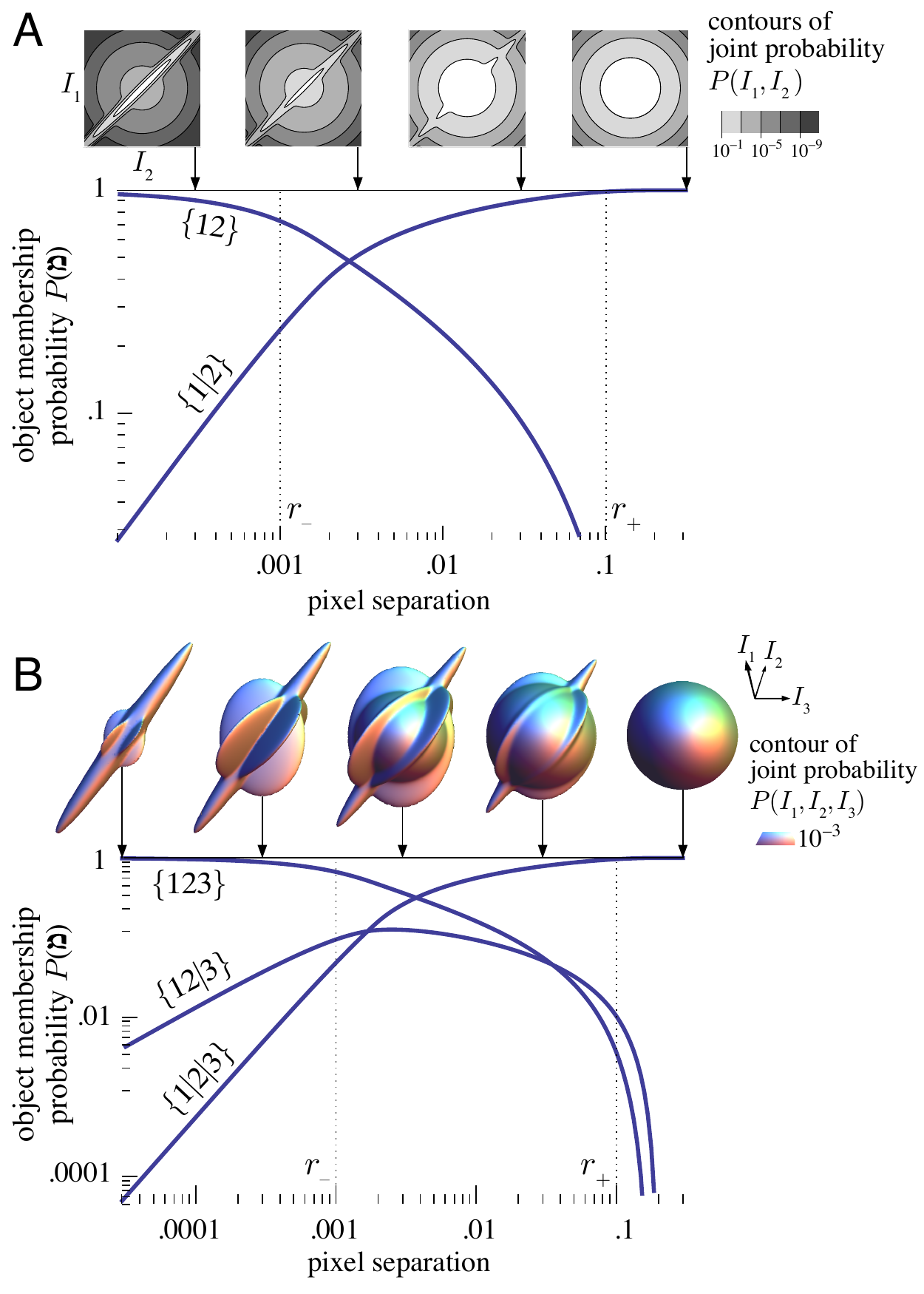}
\caption{Joint probabilities of pixel intensities, based on an ensemble of elliptical objects and gaussian-distributed object intensities with an additive gaussian white noise texture (Methods). Contour plots are shown for two pixels ({\bf A}) and three pixels arranged in an equilateral triangle ({\bf B}). These joint distributions are weighted averages of independent and correlated distributions. The weighting factors are the various object membership probabilities $P(\obj)$, which are plotted below the joint intensity distributions as a function of the distance between pixels.}
\label{contourplots}
\end{figure}

\begin{figure}[p!]
\centering
\includegraphics*[width=6.5in]{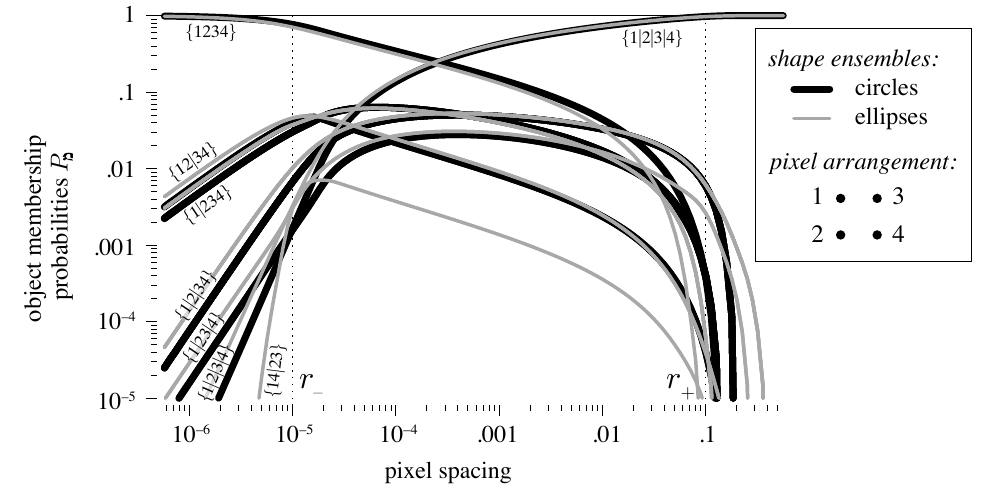}
\caption{All object membership probabilities $P(\obj)$ as a function of the spacing between the pixels, for four pixels arranged in a square (right). Two different shape ensembles are shown, circles and ellipses, with sizes given by $P(r)\propto r^{-3}$ for $r$ ranging between the limits $r_-$ and $r_+$ (dashed lines). Curves are labeled by their object membership functions. Symmetrically permuted membership functions have identical curves. For circles, the configuration $\{14|23\}$ is impossible, but otherwise the curves for circles and ellipses are remarkably similar across all pixel spacings, because both shape ensembles have similar local properties (extended edges) and global structure (convex shapes with the same size distribution). Since objects have sharp edges that closely spaced pixels rarely straddle, nearby pixels almost always fall into the same object, with $P({\{1234\}})\approx 1$. When pixel spacing exceeds the largest object dimension, no two pixels can fall into the same object, so the only membership function allowed is $\obj=\{1|2|3|4\}$. With pixel spacings between these extremes, many more object membership probabilities take on nonzero values.}
\label{Pmems}
\end{figure}

\begin{figure}[p!]
\centering
\includegraphics*[width=7in]{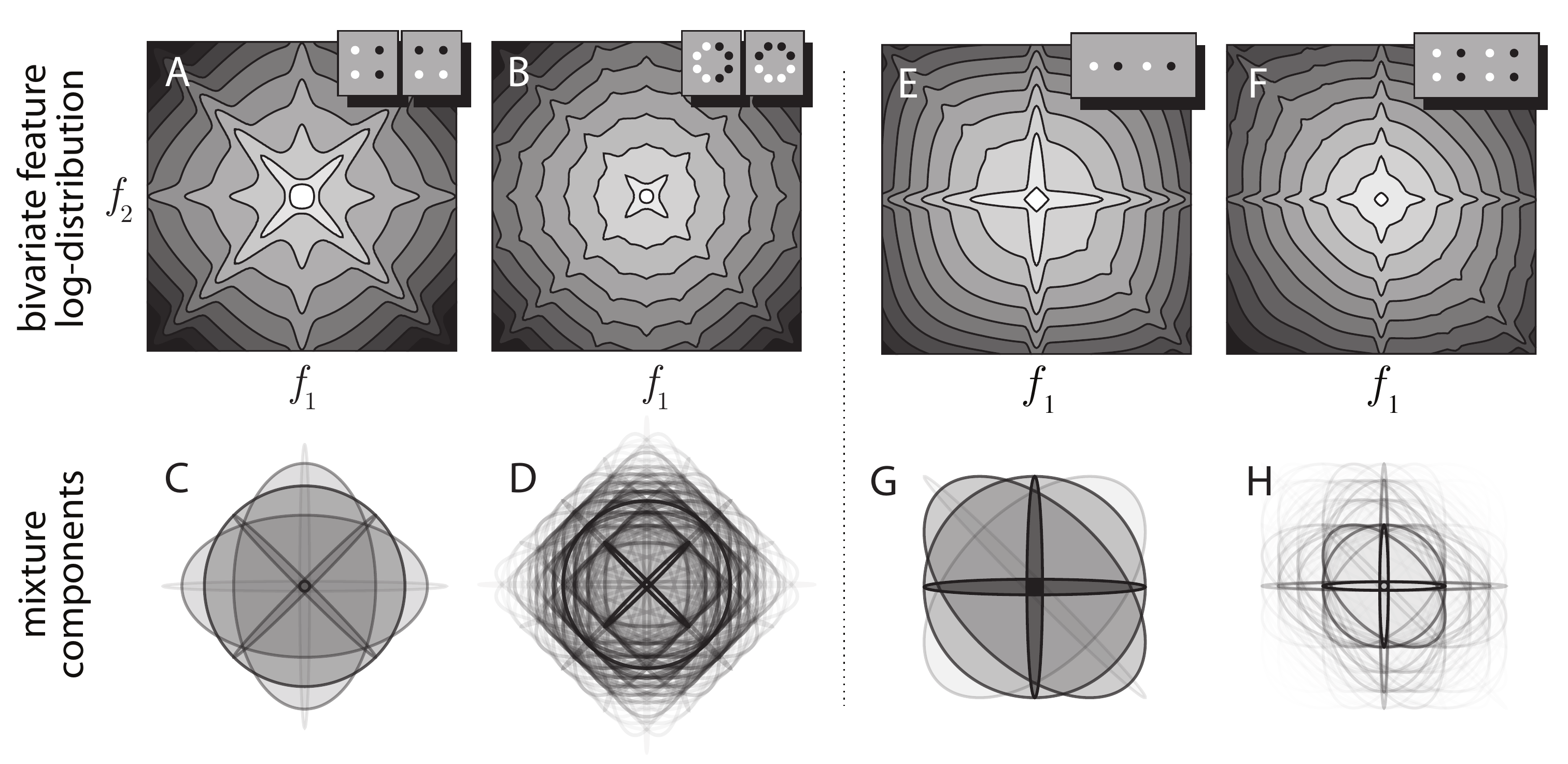}
\caption{Mixture distributions and mixture components of sparsely sampled Haar wavelet features, calculated exactly for dead leaves images. ({\bf A},{\bf B}) Contours of the log-probabilities $\log{P(f_1,f_2)}$ for colocalized, orthogonal wavelets $f_1$ and $f_2$, sampled with four or eight points per feature (insets). ({\bf C},{\bf D}) Elliptical contours of jointly gaussian mixture components $P(f_1,f_2|\obj)$, shaded according to their weight $P(\obj)$. The mixture distributions already have rounded diamond contours formed from weakly correlated components, as well as some strongly correlated and anti-correlated components which appear at all angles with dense sampling (Figure \ref{jointwavelet}B). ({\bf E}--{\bf H}) Joint log-probabilities and mixture components for nearby, parallel Haar wavelets, plotted as in A--D. The anticorrelation and `pillow' shape of these distributions are already visible with sparse sampling of the features.}
\label{jointwaveletsupplement}
\end{figure}

\begin{figure}[p!]
\centering
\includegraphics*[width=4in]{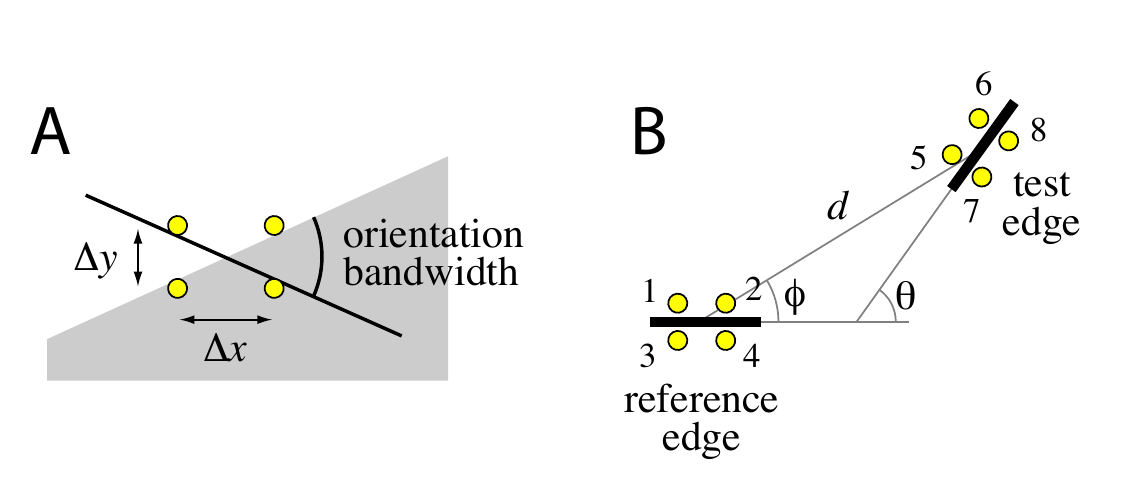}
\caption{Detailed geometry for Figure \ref{GeislerPlots}. ({\bf A}) An edge exists when an object splits four pixels into two pairs. Pixels arranged in a rectangle with an aspect ratio of $\Delta x/\Delta y=2.75$ permit a range of edges with a $40^\circ$ orientation bandwidth as used in \cite{Geisler:2001p808}. ({\bf B}) Pairs of edges thus defined are related by three parameters: distance $d$, orientation difference $\theta$, and relative direction $\phi$. }
\label{geislerplotSI}
\end{figure}

\begin{figure}[p!]
\centering
\includegraphics*[width=4.5in]{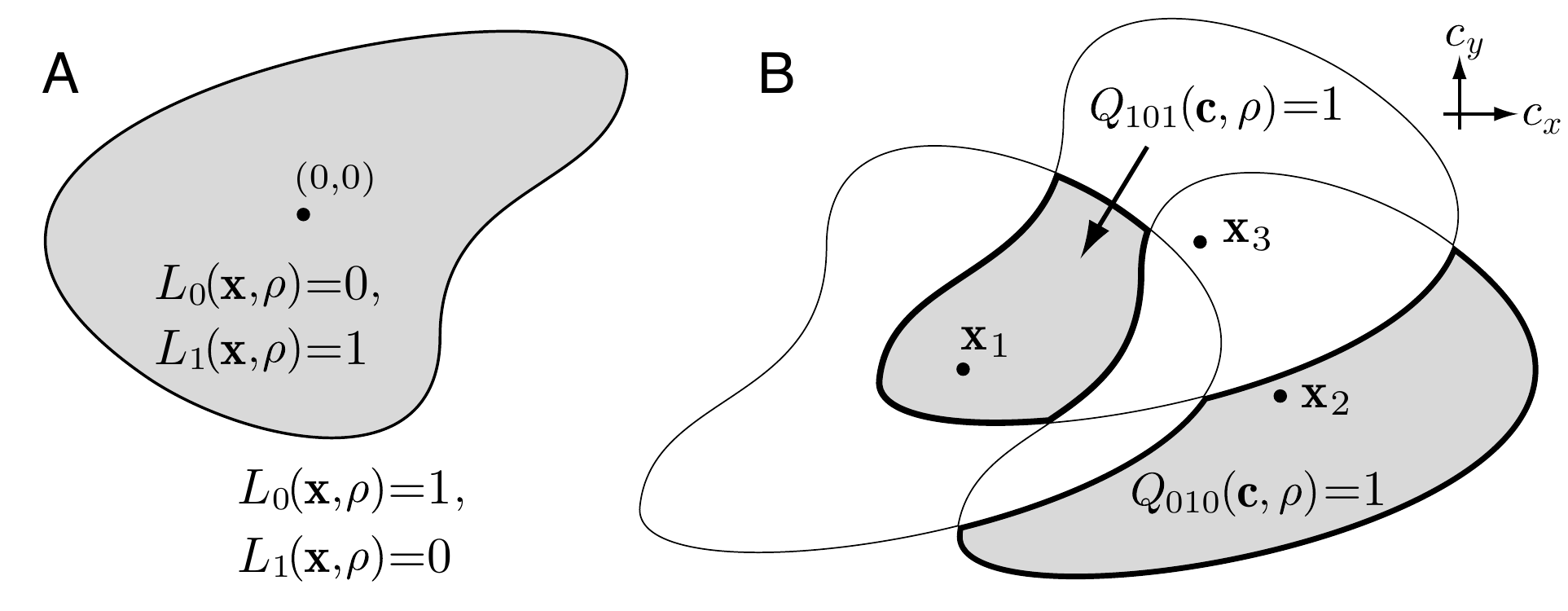}
\caption{Diagrams for illustrating inclusion probabilities $Q_{\bbsigma(\obj,n)}$. ({\bf A}) A `leaf' function showing the shape of an object. $L_1(\bx,\rho)=1$ at points $\bx$ that are inside an object of shape parameter $\rho$, and $L_0(\bx,\rho)=1$ at points outside it. Here the shape parameter $\rho$ specifies a smooth irregular object. ({\bf B}) Example indicator functions $Q_{\bbsigma(\obj,n)}(\bc,\rho)$ identify locations $\bc$ where an object could be placed to enclose all pixels $\bx_i\in\obj_n$ and exclude the rest. Rotated copies of the object shape surround each point $\bx_i$, designating the locations $\bc$ where that object will enclose $\bx_i$. An arrow points to the shaded region where an object could be placed to enclose both $\bx_1$ and $\bx_3$ but not $\bx_2$, whose area is $Q_{101}(\rho)$. The other shaded region indicates locations where an object would enclose only $\bx_2$, whose area is $Q_{010}(\rho)$. Note that this diagram represents possible locations $\bc$ of a {\em single} object, not three objects!}
\label{QsFigure}
\end{figure}

\begin{figure}[p!]
\centering
\includegraphics*[width=5.5in]{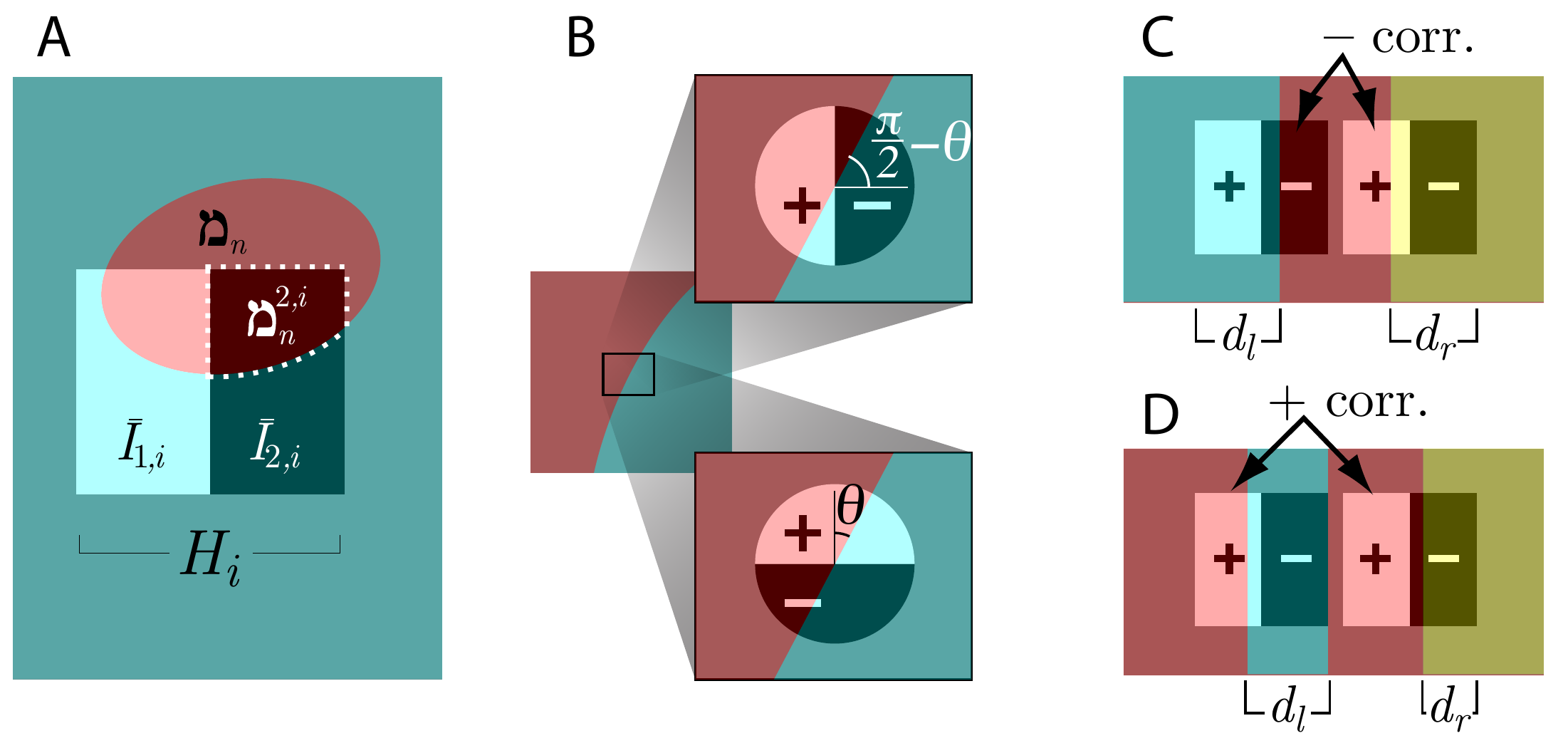}
\caption{Simplified representations of object configurations that dominate feature distributions at high amplitudes. Colors indicate objects of unspecified intensity, shading indicates weighting by Haar wavelets. ({\bf A}) A Haar wavelet $H_i$ takes a difference of intensities $\bar{I}_{1,i}$ and $\bar{I}_{2,i}$ each totalled over a finite region. The pixels $\obj_n^{2,i}$ contained both in elliptical object $\obj_n$ and in region $2$ of wavelet $i$ are outlined. ({\bf B}) Colocalized, orthogonal Haar wavelets with circular support. ({\bf C}, {\bf D}) Parallel, nearby Haar wavelets, with objects that induce negative and positive correlations, respectively. To simplify the calculations, objects differ only in their horizontal extent, and extend completely to either the left or right edge of each wavelet. The relevant variable is then the width of the overlap between the object and the wavelet filter, denoted $d_l$ and $d_r$. Compare these simplified configurations to those shown in Figures \ref{jointwavelet}C and \ref{jointwavelet}G,H.}
\label{overlapapprox}
\end{figure}

\clearpage


\section{Inclusion probabilities for an ensemble of ellipses}
\label{bigcalc}

In the main text we reported a universal recursion equation expressing object membership probabilities $P(\obj)$ in terms of some geometric factors $Q_{\bbsigma(\obj,n)}$ which depend on the shape ensemble. There we showed how these probabilities could be expressed geometrically, by first averaging $Q_{\bbsigma(\obj,n)}(\bc,\rho)$ over position $\bc$ via contour integrals, and then averaging over the shape ensemble $\rho$. Here we explain in detail how inclusion probabilities $Q_{\bbsigma(\obj,n)}$ can be calculated exactly for an ensemble of circular objects. We then use a simple transformation to generalize the result for circles to an ensemble of ellipses. When the dust settles, we will have averaged $Q_\bbsigma(\bc,\rho)$ over positions $\bc$ and shapes $\rho$ and for all binary vectors $\bbsigma$.

For an ensemble of circles, the shape parameter $\rho$ is just a radius $r$, which we draw from a scale-invariant size distribution $P(r)\propto r^{-3}$. Circular contours are easy to express analytically. However, as described in the main text, the integrals of $Q_{\bbsigma}(\bc,r)$ over both the contours and size ensemble are more difficult because they must be done piecewise. We do this in two steps. First, we evaluate the general form of the indefinite integrals at the endpoints of the piecewise intervals. Second, we describe an algorithm that synthesizes these isolated contributions into the complete piecewise integral, yielding the desired $Q_{\bbsigma}$.

\subsection{Parameterizing circular contours}

\begin{figure}[htbp]
\centering
\includegraphics*[width=7in]{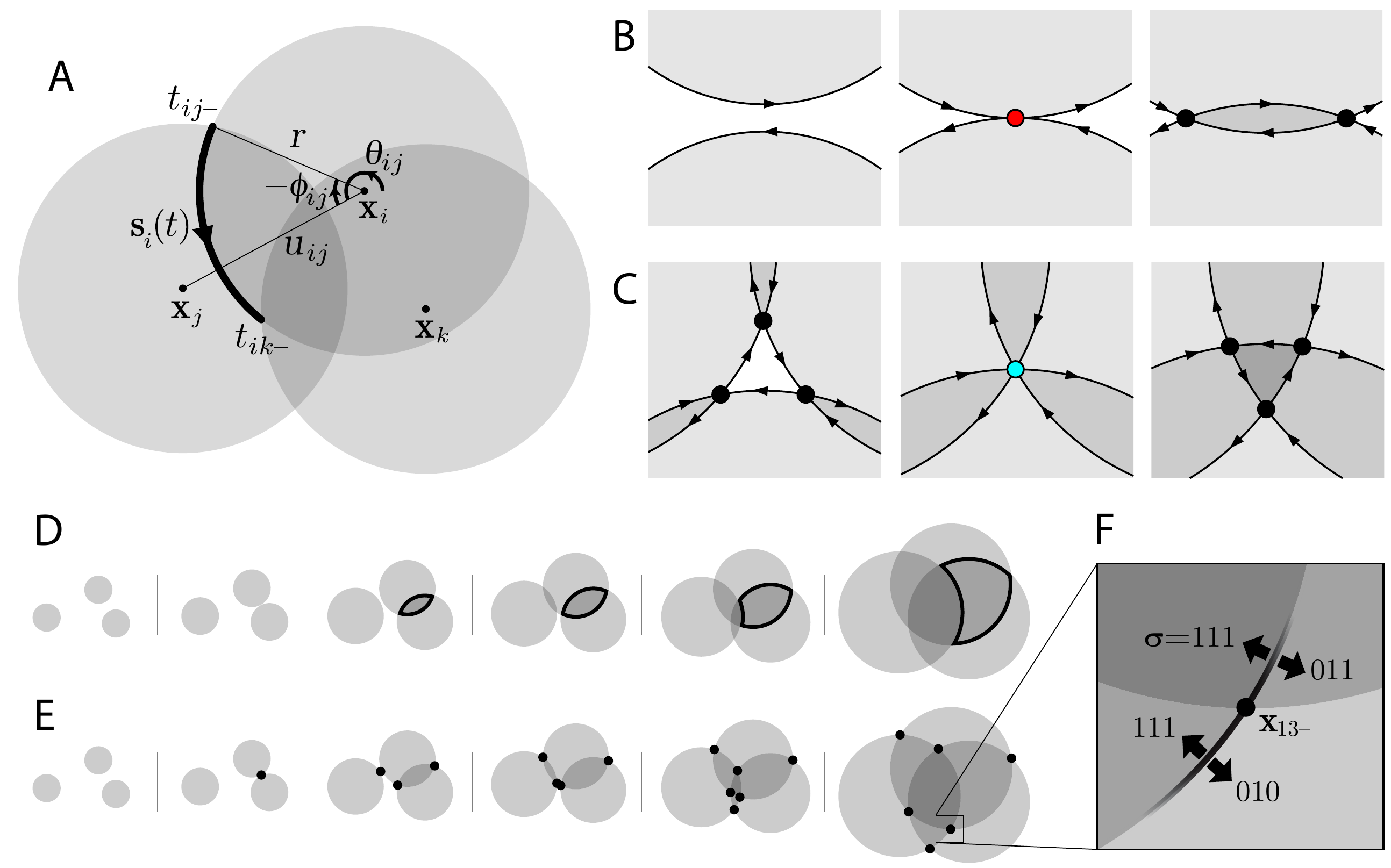}
\caption{({\bf A}) Diagram depicting the quantities needed to calculate inclusion probabilities $Q_{\bbsigma(\obj,n)}(r)$. The different regions of constant $Q_{\bbsigma(\obj,n)}(\bc,r)$ for fixed $r$ are bounded by circular arcs centered on the pixels $\bx$. Highlighted is one particular arc $\bs_i(t)$ centered on point $\bx_i$. This arc is bounded by $t_{ij-}$ and $t_{ik-}$, two angles at which other circles intersect. Centers $\bx_i$ and $\bx_j$ are separated by the distance $u_{ij}$ and angle $\theta_{ij}$. The location at which the corresponding circles intersect deviates from the line connecting the centers by angle $-\phi_{ij}$, so that $t_{ij-}=\theta_{ij}-\phi_{ij}$. ({\bf B}) Illustration of how the contours around regions with constant $Q_{\bbsigma(\obj,n)}(\bc,r)$ change shape as $r$ increases (from left to right). Two regions in $\bc$-space first touch when $r$ equals half the distance between two pixels $\bx_i$ and $\bx_j$, a critical radius $r^*_{ij}$ we call a `kissing point' (center panel). As $r$ increases further a new contour of reversed orientation is created, bounding a region within which an object of radius $r$ can enclose both pixels. ({\bf C}) Similarly, a `triple intersection' always exists for a particular $r^*_{ijk}$, the circumradius, at which any three non-collinear pixels $\bx_i$, $\bx_j$ and $\bx_k$ are equidistant from a fourth point called the circumcenter (center panel). As $r$ crosses this critical radius, the existing contour connecting the three intersection points changes orientation, and the enclosed region is associated with a different $Q_{\bbsigma(\obj,n)}$. ({\bf D},{\bf E}) Illustrations of two strategies for integrating $Q_{\bbsigma}(r)$ over $r$: choose only one region at a time, and track its contour as $r$ varies (D); or track all contour endpoints over $r$ and add their contributions to all appropriate regions (E). We use the latter strategy. Arrows in panel {\bf F} depict the four regions $\bbsigma$ receiving identical contributions (up to a sign) from the contours along $\bs_1(t)$ that terminate at the intersection point $\bx_{13-}$.}
\label{circlegeometries}
\end{figure}

Equation \ref{divergencetheorem} related the positional average $Q_{\bbsigma(\obj,n)}(\rho)$ to the total area of the region where $Q_{\bbsigma(\obj,n)}(\bc,\rho)=1$, and thence to a contour integral. In this section we evaluate this contour integral for circles with fixed radius, so that $\rho=r$. It is helpful to change from the generic notation used in Section \ref{inclusionprobs} to a notation which is specific to circular objects. As shown in Figure \ref{circlegeometries}A, the boundaries of regions with constant $Q_{\bbsigma(\obj,n)}(\bc,r)$ are all circular arcs centered on some point $\bx_i$,
\begin{equation*}
\bs_i(t)=r\be_t+\bx_i
\end{equation*}
with a unit vector defined as $\be_t\equiv(\cos{t}, \sin{t})$. Each arc terminates at angles $t$ of the form
\begin{equation*}
t_{ij\pm}=\theta_{ij}\pm\phi_{ij}=\tan^{-1}{(\bx_i-\bx_j)}\pm\cos^{-1}{(u_{ij}/2r)}
\end{equation*}
where $\theta_{ij}$ is the angle of the line connecting the circle centers, and $\pm\phi_{ij}$ are the angles that the intersection points make with that line (Figure \ref{circlegeometries}A). $\theta_{ij}$ is independent of $r$, whereas $\phi_{ij}$ depends on the ratio of $r$ to the distance $u_{ij}=|\bx_i-\bx_j|$ between the circles as $\phi_{ij}=\cos^{-1}{(u_{ij}/2r)}$. 

\subsection{Contour integration}

Since the unit normal vectors are simply $\hat{\bn}(t)=\be_t$ and the arc length is $ds=|\dot{\bs}(t)|dt=r\,dt$, we can now easily perform the contour integral (Equation \ref{contourintegral}) over each arc analytically.
\begin{equation*}
A_m=\frac{1}{2}\int_{t'_m}^{t''_m}{\bs_m(t)\cdot \hat{\bn}(t)}\,ds=\frac{1}{2}\int_{t_{ij\pm}}^{t_{ik\pm}}\left(r^2+r{\bf x}_{i}\cdot \be_t\right)dt=a_{ik\pm}(r)-a_{ij\pm}(r)
\end{equation*}
where we have defined
\begin{align}
a_{ij\pm}(r)&=\frac{1}{2}\left(r^2t_{ij\pm}+r\bx_i\cdot\be_{t_{ij\pm}-\frac{\pi}{2}}\right)\notag\\
&=\frac{r^2\theta_{ij}}{2}\pm \frac{r^2}{2}\cos^{-1}{\frac{u_{ij}}{2r}}+\frac{u_{ij}}{4}\bx_i\cdot\be_{\theta_{ij}-\frac{\pi}{2}}\pm \frac{r}{2}\bx_i\cdot\be_{\theta_{ij}}\sqrt{1-\frac{u_{ij}^2}{4r^2}}
\label{aendpoint}
\end{align}
For $r$ smaller than the distances between pixels, the circular arcs do not intersect and are thus complete circles with total area of $\pi r^2$, as expected.

Finally we can obtain the total area $Q_{\bbsigma(\obj,n)}(r)$ by adding up the relevant $a_{ij\pm}(r)$ appropriately. We defer discussion of this step to Section \ref{algorithm}.

\subsection{Indefinite integral over radius}

Next we have to average these quantities over the distribution of $r$. To achieve scale invariance in generated images, the distribution of object radii $P(r)$ should be proportional to $r^{-3}$ \cite{Lee:2001p1}. Some deviation from this scaling behavior is required to prevent images from degenerating with high probability into white noise or uniform coloring \cite{Lee:2001p1,Bordenave:2006p2952}. Here we choose to set upper and lower size cutoffs $r\in[r_-,r_+]$ to satisfy this constraint, so that
$$
P(r)=\begin{cases}\frac{1}{z}r^{-3} & r\in[r_-,r_+]\\0 & \rm{otherwise}\end{cases}
$$
with $z=\tfrac{1}{2}(r_-^{-2}-r_+^{-2})$.

The $r$-dependence of $a_{ij\pm}(r)$ in Equation \ref{aendpoint} takes the forms $r\sqrt{1-\left(u/2r\right)^2}$, $r^2\cos^{-1}{(u/2r)}$, and $r^2$. Each of these terms must be averaged over $P(r)$. The first average can be solved analytically as
\begin{align*}
\int\!P(r)&r\sqrt{1-\left(u/2r\right)^2}dr=\frac{1}{z}\int\!r^{-2}\sqrt{1-\left(u/2r\right)^2}=-\frac{1}{u}\left[\frac{1}{2}\sin{2\phi}+\frac{\pi}{2}-\phi\right]
\end{align*}

The average of the second term can be found in tables of integrals, and involves a special function, the dilogarithm $\Li_2(z)$.
\begin{align*}
\int\!P(r)&r^2\cos^{-1}{\!\frac{u}{2r}}\,dr=\frac{1}{z}\int\!r^{-1}\cos^{-1}{\!\frac{u}{2r}}\,dr=\frac{i}{2}\phi^2-\phi\log{\left(1+e^{2i\phi}\right)}+\frac{i}{2}\,\Li_2{\left(-e^{2i\phi}\right)}
\end{align*}
For $u<2r$ (required for the two relevant circles to intersect), the imaginary component is constant and therefore cancels in any real definite integral. We can therefore take just the real component without influencing the result.
$$
\int\!P(r)r^2\cos^{-1}{\!\frac{u}{2r}}\,dr=-\phi\log{\frac{u}{2r}}-\frac{1}{2}\Re{\left[i\,\Li_2\left(-e^{2i\phi}\right)\right]}
$$
The imaginary part of the dilogarithm evaluated on the complex unit circle is related to another special function known as Clausen's integral, for which optimized numerical routines have been written [48].
$$
\Re\left[i\,{\rm Li}_2\left(-e^{2i\phi}\right)\right]={\rm Cl}_2\left(-2\phi-\pi\right)
$$
The remaining terms in $a_{ij\pm}(r)$ are elementary to integrate: $\int P(r)r^2dr=\frac{1}{z}\log{r}$ and $\int P(r)dr=-\frac{1}{2z}r^{-2}$.

Combining all these pieces with their correct coefficients, we obtain the indefinite integral for the size average of $a_{ij\pm}(r)$.
\begin{align}
b_{ij\pm}(r)\equiv\int\!dr\,P(r)a_{ij\pm}(r)
=&-\frac{1}{2z}\theta\log{r}+\frac{u}{8zr^2}\bx_{i}\cdot\be_{\theta-\frac{\pi}{2}}\notag\\
&\pm\frac{1}{2zu}\left(\frac{1}{2}\sin{2\phi}+\frac{\pi}{2}-\phi\right)\bx_{i}\cdot\be_{\theta}\notag\\
&\pm\frac{1}{2z}\phi\log{\left(\frac{u}{r}\right)}\pm\frac{1}{4z}{\rm Cl}_2{\left(-2\phi-\pi\right)}
\label{VertexEquation}
\end{align}

\subsection{Identifying piecewise smooth intervals over radius}

The definite integral over $r$ must be performed piecewise because its integration contours may change at certain critical radii $r^*$. Generically, there are two types of critical radii, depicted in Figure \ref{circlegeometries}B,C: `kissing points' where $r^*_{ij}$ is half the distance $u_{ij}$ between a pair of points $\bx_i$ and $\bx_j$, so that their corresponding circles just touch; and `triple intersections' where $r^*_{ijk}$ equals the circumradius of three points $\bx_i$, $\bx_j$ and $\bx_k$, so that the three corresponding circles all meet. For three points separated by distances $u_{ij}$, $u_{jk}$, and $u_{ki}$, and semiperimeter $s=\frac{1}{2}(u_{ij}+u_{jk}+u_{ki})$, the circumradius is
$$r^*_{ijk}=u_{ij}u_{jk}u_{ki}/4\sqrt{s(s-u_{ij})(s-u_{jk})(s-u_{ki})}$$
If the pixel locations have extra symmetries, e.g. lie on a lattice, then several critical radii $r^*$ may coincide. In this case each $r^*$ can be treated sequentially without changing the result, as if perturbing each $r^*$ infinitesimally: $b_{ij\pm}(r'')-b_{ij\pm}(r')$ contributes zero in the limit $r''-r'\to 0$ when there are no intervening critical radii.

\subsection{Mapping piecewise integrals onto appropriate $Q_\bbsigma$}
\label{algorithm}

Now we must calculate $Q_\bbsigma$ by adding up the definite integral $b_{ij\pm}(r)$ evaluated at the appropriate critical radii $r^*$ and the relevant triples $(i,j,\pm)$. Consider two strategies for this. First, one could choose one particular $\bbsigma$, and track how the cusps of $Q_\bbsigma(\bc,r)$'s boundary appear, change, and disappear as a function of $r$, and then add up the appropriate contributions from Equation \ref{VertexEquation} (Figure \ref{circlegeometries}D). One would then repeat this procedure for every possible $\bbsigma$. Second, one could choose a particular intersection point $\bx_{ij\pm}$ between two objects, track how it is associated with different regions as a function of $r$, and add its contribution to the various appropriate $Q_\bbsigma$. By iterating through all intersection points, eventually all contributions to all $Q_\bbsigma$ are computed (Figure \ref{circlegeometries}E). This latter strategy is easier because the behavior of the intersection points is simpler to track than the various (possibly unconnected) regions where $Q_\bbsigma(\bc,r)=1$. This is the approach we describe below.

To compute the definite integral corresponding to Equation \ref{VertexEquation} above, we must therefore associate each integrand $a_{ij\pm}(r)$ with boolean vectors $\bbsigma$ designating the correct targets $Q_\bbsigma$ for each interval of $r$. The region geometry, and thus these desired associations, change only at critical radii; between critical radii the associations are constant. By construction, $a_{ij\pm}(r)$ (Equation \ref{aendpoint}) is the result of a contour integral terminating at an intersection between circles centered on $\bx_i$ and $\bx_j$ (Figure \ref{circlegeometries}A). We label this intersection point by $\bx_{ij\pm}=\bx_i+r\be_{\theta_{ij}\pm\phi_{ij}}$. Contour integrals terminating at this point contribute to every one of the four regions that touch $\bx_{ij\pm}$, i.e. the $\bbsigma$ involving all four allowed combinations of its elements $\sigma_i\in\{0,1\}$ and $\sigma_j\in\{0,1\}$ (Figure \ref{circlegeometries}F). The point $\bx_{ij\pm}$ is not on the boundary of any circles centered on other pixels $\bx_\ell$, since otherwise there would be a critical radius within the selected $r$ interval. $\bx_{ij\pm}$ is thus either strictly inside or strictly outside a circle of radius $r$ for all $\ell\neq i,j$. We can now specify all elements of $\bbsigma$ as
\begin{equation*}
\sigma_\ell=\begin{cases}
L_1(\bx_{ij\pm}-\bx_\ell,r) & \ell\neq i,j\\
\text{0 or 1} & \ell=i,j
\end{cases}
\end{equation*}
where $L_1(\bx_{ij\pm}-\bx_\ell,r)$ is the leaf function from Methods Section \ref{inclusionprobs}. This relation identifies the appropriate targets $Q_\bbsigma$ for the $b_{ij\pm}$ of Equation \ref{VertexEquation}.

To identify the {\it signs} $c_{ij\pm}$ with which $b_{ij\pm}$ contribute to the target $Q_\bbsigma$, it helps to go back and compute the signs that $a_{ij\pm}(r)$ contribute to the target area $Q_\bbsigma(r)$. These signs depend on the geometry of the region contours. Consider how the region boundaries change their geometry as $r$ increases from $r_-$ to $r_+$. A contour around an object boundary is counterclockwise initially, i.e. before the contour intersects any other object boundaries. As $r$ increases past a kissing point $r^*_{ij}$, a pair of intersection points $\bx_{ij\pm}$ is created along with a new region with clockwise orientation (Figure \ref{circlegeometries}B). Note that the contours around the object centered on $\bx_i$ initially {\it converge} at an intersection $\bx_{ij-}$ and {\it diverge} at $\bx_{ij+}$. In other words, intersections $\bx_{ij-}$ are initially endpoints of the contours along $\bs_i(t)$ that contribute $+a_{ij\pm}$ to the contour integral (Equation \ref{contourintegral}), and $\bx_{ij+}$ are initially starting points that contribute $-a_{ij\pm}$. However, as $r$ increases past each triple-intersection $r^*_{ijk}$ for $k\neq i,j$, another circle centered on $\bx_k$ encloses the intersection point. The orientations of the contours at $\bx_{ij\pm}$ then reverse (Figure \ref{circlegeometries}C), and the sign that each $a_{ij\pm}$ contributes also reverses. Thus the overall convergence for paths at an intersection point is: converging for $-$, diverging for $+$, and reversed by the number of circles enclosing the point. Mathematically, we can write the desired sign as
\begin{equation*}
c_{ij\pm}(r)=\mp(-1)^{\sum_{\ell\neq i,j} L_1(\bx_\ell-\bx_{ij\pm},r)}
\end{equation*}
Note that $c_{ij\pm}(r)$ does not vary between critical radii $r^*$, so we may use its value anywhere within the integration interval. Finally, when we integrate $a_{ij\pm}(r)$ over $r'<r<r''$, the value of the indefinite integral $b_{ij\pm}$ at $r'$ is subtracted from the value at $r''$. Thus, for each interval between critical radii we add
\begin{equation*}
\Delta Q_\bbsigma(r',r'')=c_{ij\pm}(\tfrac{r'+r''}{2})\cdot\big(b_{ij\pm}(r'')-b_{ij\pm}(r')\big)
\end{equation*}
to the appropriate $Q_\bbsigma$.

There is one remaining subtlety in adding up the contributions to $Q_\bbsigma$. In the first term of $b_{ij\pm}$ there is an ambiguity of $2\pi$ in what angle is subtended by a given arc, which cannot be resolved by local properties of the arc endpoints alone. We remedy this by computing $\Delta Q_\bbsigma(r',r'')$ modulo $\tfrac{\pi}{z}\log{r''/r'}$, which is the maximum possible contribution an area can make between $r'$ and $r''$. This guarantees that we update $Q_\bbsigma$ with the unique definite integral over $r'<r<r''$ that lies between 0 and this maximum.

\subsection{Summary of the algorithm for calculating $Q_\bbsigma$}
\label{Algorithm}

This completes the mathematics necessary to calculate the $Q_\bbsigma$. To summarize, we present the method in algorithmic form.

\begin{enumerate}
	\item Initialize all $Q_\bbsigma$ to zero.
	\item Add $\int_{r_-}^{r^*_i} dr\,P(r)\pi r^2=\frac{\pi}{z}\log{\frac{r_i^*}{r_-}}$ to $Q_{{\bf \delta}_i}$ for each circle, where $r_i^*=\min_{j\neq i}{r^*_{ij}}$ is the first kissing point for that circle and ${\bf \delta}_i$ is a vector of zeros with a 1 at index $i$. This is the area accumulated in $Q_{{\bf \delta}_i}$ before any other circles were touched.
	\item Sort all critical radii $r^*_{ij}$ and $r^*_{ijk}$ within the integration bounds $r_-$ and $r_+$.
	\item For each interval $r'<r<r''$ bounded by sequential critical radii:

	\begin{enumerate}
		\item For each existing intersection point $\bx_{ij\pm}$:
	
		\begin{enumerate}
			\item Calculate the region indicators $\bbsigma$ to which the point $\bx_{ij\pm}$ contributes
			\item Add $\Delta Q_\bbsigma(r',r'')$ modulo $\frac{\pi}{z}\log{\frac{r''}{r'}}$ to $Q_\bbsigma$
		\end{enumerate}
	\end{enumerate}
	\item Set $Q_{\bf 0}=1-\sum_{\bbsigma\neq \bf 0} Q_\bbsigma$.
\end{enumerate}

Once the $Q_{\bbsigma(\obj,n)}$ are calculated for all object membership functions $\obj$, then $Q_{\bbsigma(\obj_{\setminus n},k)}$ must be calculated for the reduced $\obj_{\setminus n}$ used in the recursion. For efficiency, this can be accomplished by marginalizing $Q_\bbsigma$ over the appropriate indices $\sigma_i$, rather than recalculating it with a smaller set of pixels.

Note that with a different size ensemble $P(r)$, the expression for $b_{ij\pm}$ would change, but the procedure for combining them to obtain the $Q_\bbsigma$ would be the same.

\subsection{Converting from circles to ellipses}
\label{ellipses}

It is straightforward to transform our calculation of $Q_\bbsigma$ for circles into a result for ellipses of equal area but eccentricity $\epsilon$ and orientation $\psi$. All distances are effectively scaled by $\sqrt{\epsilon}$ in the direction of $\be_{\psi}$ and $1/\sqrt{\epsilon}$ in the orthogonal direction. This is equivalent to transforming the pixel locations $\bx_i\to\bx'_i$ as
$$\bx'_i=\frac{1}{\sqrt{\epsilon}}\left(\begin{array}{cc} \epsilon\cos^2{\psi}+\sin^2{\psi} & (\epsilon-1)\cos{\psi}\sin{\psi} \\ (\epsilon-1)\cos{\psi}\sin{\psi} & \cos^2{\psi}+ \epsilon\sin^2{\psi} \end{array}\right)\cdot\bx_i$$
and recomputing the $Q_{\bbsigma}$ with these $\bx'_i(\epsilon,\psi)$.

Unfortunately we cannot analytically integrate $b_{ij\pm}(r^*)$ as a function of eccentricity $\epsilon$ or angle $\psi$, because the dependence on the points $\bx'_i(\epsilon,\psi)$ already involves special functions. Instead, to obtain the average over possible ellipses we use a discrete ensemble of eccentricities and angles and sum over them as
$$\left\langle Q_\bbsigma(\bc,r,\epsilon,\psi)\right\rangle_{\bc,r,\epsilon,\psi}=\sum_{\epsilon,\psi}Q_\bbsigma(\epsilon,\psi)P(\epsilon)P(\psi)$$

More generally, when the integral cannot be expressed analytically using easily computable functions, one may specify the ensemble $P(\rho)$ by a discrete number of allowed shapes, and compute the ensemble average as a sum rather than as an integral.

The result of these calculations are concrete numbers for the inclusion probabilities $Q_\bbsigma$, which can then be substituted into Equations \ref{factordistribution}, \ref{mixturedistribution}, and \ref{recursion} to calculate the object membership probabilities and joint distributions of pixel intensities and image features.

\begin{table}[htbp]
\centering
\begin{tabular}{| l l |}
\hline
$\obj$ & Hebrew letter {\it mem}: an object membership function\\
$|\obj|$ & number of distinct objects in $\obj$ \\
$\obj_n$ & set of pixels contained in $n$th object\\
$\obj_{\setminus n}$ & object membership function with $n$th object removed \\
$P_\obj$ & probability that pixels are divided according to $\obj$ \\
\hline
$\bx$ & locations of all $N$ selected pixels \\
$\bx_i$ & location of $i$th pixel\\
$\bI$ & vector of all $N$ pixel values \\
$I_i$ & pixel value at point $\bx_i$\\
$\bI_{\obj_n}$ & vector of all pixel values in $n$th object \\
\hline
$\bbsigma(\obj,n)$ & boolean vector indicating pixels in $n$th object\\
$Q_{\bbsigma(\obj,0)}$ & probability that an isolated object includes no selected pixels\\
$Q_{\bbsigma(\obj,n)}$ & probability that an isolated object includes only pixels $\obj_n$\\
$Q_{\bbsigma(\obj,n)}(\rho)$ & as above, but given the object shape $\rho$\\
$Q_{\bbsigma(\obj,n)}(\bc,\rho)$ & as above, but also given the object position $\bc$\\
\hline
$\bx_{ij\pm}$ & location of two intersections between objects centered on $\bx_i$ and $\bx_j$\\
$u_{ij}$ & distance between $\bx_i$ and $\bx_j$\\
$\theta_{ij}$ & angle of the vector $\bx_j-\bx_i$\\
$\phi_{ij}$ & absolute value of angle between intersection points and line connecting $\bx_i$ and $\bx_j$\\
$t_{ij\pm}$ & angle of $\bx_{ij\pm}-\bx_i$\\
$\be_\theta$ & unit vector $(\cos{\theta},\sin{\theta})$\\
\hline
$\bs_i(t)$ & contour around object centered on $\bx_i$\\
$a_{ij\pm}(r)$ & indefinite integral over $t$ along contour $\bs_i(t)$ evaluated at $t_{ij\pm}$ with fixed $r$\\
$b_{ij\pm}(r)$ & indefinite integral of $a_{ij\pm}(r)$ over $r$\\
$c_{ij\pm}(r)$ & sign indicating whether contour $\bs_i(t)$ starts or ends at $t=t_{ij\pm}$\\
\hline
$r^*$ & critical radius at which regions of constant $Q_{\bbsigma(\obj,n)}(\bc,\rho)$ change structure\\
$r^*_{ij}$ & radius of kissing point for circles on $\bx_i$ and $\bx_j$\\
$r^*_{ijk}$ & radius of triple intersection for circles on $\bx_i$, $\bx_j$ and $\bx_k$\\
\hline
$\epsilon$ & ellipse eccentricity\\
$\psi$ & orientation of major axis of ellipse\\
$\bx'_i(\epsilon,\psi)$ & transformed pixel location\\
\hline
\end{tabular}
\caption{Glossary of symbols used}
\label{notation}
\end{table}



\section*{References for Supporting Information}
[48]\ \ MacLeod A (1996) Algorithm 757, miscfun: A software package to compute uncommon special functions. ACM Transactions on Mathematical Software 22: 288--301.

\end{document}